\documentclass[floatfix,nofootinbib,twocolumn,preprintnumbers,superscriptaddress,notitlepage]{revtex4-1}

\usepackage{times}
\usepackage{xcolor}
\usepackage{amsfonts}
\usepackage{amsmath}
\usepackage{amssymb}
\usepackage{bm}
\usepackage{dcolumn}
\usepackage{enumerate}
\usepackage{epsfig}
\usepackage{graphicx}
\usepackage{graphics}
\usepackage[latin1]{inputenc}
\usepackage{latexsym}
\usepackage{rotating}
\usepackage{hyperref}
\usepackage[caption=false]{subfig}
\usepackage[normalem]{ulem}
\usepackage[columnwise]{lineno}

\newcommand{\<}{\begin{equation}}
\newcommand{\?}{\end{equation}}

\renewcommand{\arraystretch}{2}

\definecolor{mgreen}{rgb}{0.1,0.7,0.1}

\begin{document}

\title{Inferring spin tilts of binary black holes at formation with plus-era gravitational wave detectors}

\author{Sumeet Kulkarni}
\affiliation{Department of Physics and Astronomy, The University of Mississippi, University, Mississippi 38677, USA}
\author{Nathan~K.~Johnson-McDaniel}
\affiliation{Department of Physics and Astronomy, The University of Mississippi, University, Mississippi 38677, USA}
\author{Khun Sang Phukon}
\affiliation{Nikhef - National Institute for Subatomic Physics, Science Park, 1098 XG Amsterdam, The Netherlands}
\affiliation{Institute for High-Energy Physics, University of Amsterdam, Science Park, 1098 XG Amsterdam, The Netherlands}
\affiliation{Institute for Gravitational and Subatomic Physics, Utrecht University, Princetonplein 1, 3584 CC Utrecht, The Netherlands}
\affiliation{School of Physics and Astronomy and Institute for Gravitational Wave Astronomy,\\University of Birmingham, Edgbaston, Birmingham, B15 9TT, United Kingdom}
\author{N. V. Krishnendu}
\affiliation{Max Planck Institute for Gravitational Physics (Albert Einstein Institute), Callinstr.~38, D-30167 Hannover, Germany}
\affiliation{Leibniz Universit{\"a}t Hannover, D-30167 Hannover, Germany}
\affiliation{International Centre for Theoretical Sciences (ICTS), Survey No. 151, Shivakote, Hesaraghatta, Uttarahalli Hobli, Bengaluru, 560089}
\author{Anuradha Gupta}
\affiliation{Department of Physics and Astronomy, The University of Mississippi, University, Mississippi 38677, USA}

\date{\today}

\begin{abstract}
The spin orientations of spinning binary black hole (BBH) mergers detected by ground-based gravitational wave detectors such as LIGO and Virgo can provide important clues about the formation of such binaries. 
However, these spin tilts, i.e., the angles between the spin vector of each black hole and the binary's orbital angular momentum vector, can change due to precessional effects as the black holes evolve from a large separation to their merger. The tilts inferred at a frequency in the sensitive band of the detectors by comparing the signal with theoretical waveforms can thus be significantly different from the tilts when the binary originally formed. 
These tilts at the binary's formation are well approximated in many scenarios by evolving the BBH backwards in time to a formally infinite separation. Using the tilts at infinite separation also places all binaries on an equal footing in analyzing their population properties. In this paper, we perform parameter estimation for simulated BBHs and investigate the differences between the tilts one infers directly close to merger and those obtained by evolving back to infinite separation. We select simulated observations such that their configurations show particularly large differences in their orientations close to merger and at infinity. While these differences may be buried in the statistical noise for current detections,  
we show that in future plus-era (A$+$ and Virgo$+$) detectors, they can be easily distinguished in some cases. We also consider the tilts at infinity for BBHs in various spin morphologies and at the endpoint of the up-down instability. In particular, we find that we are able to easily identify the up-down instability cases as such from the tilts at infinity.
\end{abstract}

\maketitle

\section{Introduction}
\label{sec:intro}

The detection of $\sim 90$ 
compact object mergers involving black holes and neutron stars as gravitational wave (GW) events by the Advanced LIGO~\cite{LIGOScientific:2014pky} and Advanced Virgo~\cite{VIRGO:2014yos} detectors over the course of three observing runs~\cite{LIGOScientific:2021djp} has ushered in 
the new era of GW astrophysics. A majority of these detections are mergers of binary black hole (BBH) systems. The GW signal for a given BBH carries information about the intrinsic properties of the system, such as the masses, spins, and orbital eccentricity. While at current detector sensitivities we are unable to precisely determine these properties for any given BBH, a combined catalog of these events  
makes it possible to study the population properties of BBHs, estimate their local merger rate, and provides clues as to how they could have formed~\cite{LIGOScientific:2021psn}.

For a precessing BBH system, the orientations of the spin vectors of the two black holes in the binary are parameters of considerable interest, in particular these spin vectors' misalignment with the binary's (Newtonian) orbital angular momentum, $\vec{L}$. These \textit{spin tilts}, denoted by $\theta_1$ ($\theta_2)$, are defined as the angle between $\vec{L}$ and the spin vector of the primary (secondary) black hole. The spin tilts range from $0$ to $\pi$, corresponding to orientations that range from aligned to anti-aligned with $\vec{L}$. 
In addition to these spin tilts, the relative angle, between the orbital plane angles $\phi_{1,2}$ made by the two spin vectors, $\phi_{12}:=\phi_2-\phi_1$, is also crucial to provide a complete description of the (orbit-averaged) precessional dynamics of a BBH (for an illustration of these angles, see, e.g., Fig. 1 in~\cite{Johnson-McDaniel:2023oea}). Values for all these spin angles are given in radians throughout this text, unless specified otherwise.

Well constrained spin orientations may prove to be the smoking gun for distinguishing between the two most well-known formation scenarios for BBHs~\cite{Vitale:2015tea,Rodriguez:2016vmx,Farr:2017uvj,Farr:2017gtv}.
In the \textit{isolated} formation scenario (as outlined in \cite{Mandel:2021smh}), 
component stars in a field binary collapse separately to form black holes. Mass transfer between the two components and tidal interactions both tend to align the component spins with the binary's orbital angular momentum~\cite{Gerosa:2013laa} (see also~\cite{Stevenson:2017dlk} for an overview of these mechanisms), and hence isolated BBHs are generally expected to have small tilt angles. 
However, large misalignments may still be obtained in this scenario through asymmetric natal supernova kicks in the formation of the second compact object~\cite{Kalogera:1999tq}. Alternatively, black holes in a dense stellar environment may form a binary through many-body interactions in the \textit{dynamical} formation scenario (see, e.g.,~\cite{Mandel:2021smh}).
Since these encounters are stochastic, BBHs formed through this channel are expected to have an isotropic spin distribution. BBHs can also form dynamically  when embedded in the disks of active galactic nuclei powered by the inflow of matter into the central galactic supermassive black hole (e.g.,~\cite{Mandel:2021smh}). Although the modeling of these systems has significant uncertainties, a simple calculation finds that spin-orbit interactions within the disk can drive black hole spin misalignments as large as  $\gtrsim 60^\circ$~\cite{Li:2022cul}. 
An increasing number of GW observations may make it possible to distinguish aligned and isotropic spin distributions from a population of detected BBHs. The current detections in GWTC-3 already show a stronger preference for an isotropic spin tilt distribution in the population compared to previous catalogs~\cite{LIGOScientific:2021psn}.

However, the current sensitivities of the LIGO-Virgo detectors in their third observing run (O3) do not allow one to obtain strong constraints on the tilts for individual events.
With planned improvements in the sensitivity of these detectors, this picture could change. The upcoming fourth observing run (O4) of advanced LIGO and advanced Virgo will see these detectors achieve their design sensitivities. 
This will be followed by a further set of upgrades~\cite{KAGRA:2013rdx} leading to the plus-era of ground-based detectors (A+/AdV+).\footnote{While KAGRA~\cite{KAGRA:2018plz} joined the O3 observing run and will be part of O4 and beyond, we do not consider it in this study owing to the significant uncertainties in its plus-era sensitivity (see~\cite{timeline_graphic}).}
A recent study by Knee~\emph{et al.}~\cite{Knee:2021noc} found that spin tilts could be constrained to within a few tens of degrees for loud events in plus-era detectors.

Traditionally, inferred spin tilts have been reported at a fiducial detector-frame reference frequency, $f_{\rm ref}$ ($20$~Hz for most detections in GWTC-2~\cite{LIGOScientific:2020ibl}), typically chosen to be the lowest frequency used in the data analysis. This quantity may change across different events, and also between observing runs that have different detector sensitivities. Furthermore, the detector-frame $f_{\rm ref}$ cannot be an adequate common choice for all events, since binaries with varying masses, spins, and redshifts will be at different stages of their evolution when they reach the reference frequency. This creates problems when analyzing properties such as spin orientations at a population level. A proposed solution for this is to evolve all binaries to formally infinite separation to place them on the same footing~\cite{Mould:2021xst}. Indeed, in GWTC-2.1~\cite{LIGOScientific:2021usb} and GWTC-3~\cite{LIGOScientific:2021djp}, the tilts are reported at infinite separation.
An alternative proposal is to use a fixed dimensionless frequency or time near the merger for this purpose \cite{Varma:2021csh, Varma:2021xbh}, which has the advantage of also considering the in-plane spin components, which are not well defined at infinite separation. However, we choose for our reference point to compute the \textit{tilts at infinity} since they have also been shown to be a good approximation to the tilt orientations at binary formation in many binary formation scenarios, as discussed in~\cite{Johnson-McDaniel:2021rvv}. 

The term `binary formation' here refers to the formation of the second black hole in the isolated formation channel, after which the binary's evolution is governed purely by vacuum general relativity, and thus the spin evolution we consider is applicable. Prior to this, both objects' spins can be reoriented, both towards alignment through the same mass transfer and tidal processes discussed before, as well as towards misalignment from the supernova that forms the second black hole. These effects are both discussed in, e.g., \cite{Gerosa:2018wbw}. For dynamical formation, we consider the binary's formation to be after its last significant interaction with a third body (and after the formation of the second black hole), after which it can be treated as isolated to a good approximation. All these effects prior to formation of the second black hole are accounted for in population synthesis calculations. The tilts at infinity approach thus takes BBH spin orientations closer to the output of these calculations. However, as discussed in~\cite{Johnson-McDaniel:2021rvv}, for some dynamical formation scenarios the binary will never become isolated to a good approximation when it is well separated, and thus the tilts at infinity are not a good approximation to the tilts at any part of its evolution. This is notably the case for active galactic nucleus channel, where the binary interacts with the disk for a large part of its evolution (see, e.g.,~\cite{Grobner:2020drr}), as well as channels where the merging binary remains bound to a sufficiently close third body through merger (see, e.g.,~\cite{Liu:2020gif} for examples of such formation scenarios, though we have not assessed whether these specific formation scenarios have a portion of the evolution where the tilts would be well approximated by the tilts at infinity).

In this paper, we investigate how the inferred distributions of spin tilts can differ at $f_{\rm ref}$ and infinity. We first determine the spin angles at the reference frequency that give particularly large differences in the tilts at these two points, for a set of masses and spin magnitudes. 
We then perform parameter estimation on simulated GW signals with these parameters using the anticipated A+/AdV+ sensitivity. This provides posterior probability distributions for the binaries' parameters, including their spin angles at $f_{\rm ref}$. We then evolve the posterior samples to infinite separation
to obtain the tilts at infinity. For this purpose, we use a hybrid framework that combines orbit-averaged evolution close to merger with precession-averaged evolution at large binary separation. See~\cite{Johnson-McDaniel:2021rvv} for a comprehensive description of the hybrid evolution code.

We find that entire spin tilt posterior distributions can show large variations when evolved to infinite separation for BBHs detected in the A+/AdV+ era, but the largest deviations are seen only for a subset of BBHs that have comparable masses and high spin magnitudes.
For the majority of the parameter space considered in this paper, the distributions are not easily distinguishable even when considering binaries whose spin orientations are selected to obtain large differences.

We also study the behavior of the tilts at infinity for BBHs exhibiting specific types of precessional dynamics. The first consists of binaries in the C, L$0$, and L$\pi$ precessional spin morphologies introduced in~\cite{Kesden:2014sla,Gerosa:2015tea}. Specifically, we use the results from~\cite{Johnson-McDaniel:2023oea}, which reports the capability of distinguishing these morphologies through parameter estimation in plus-era detections. 
In our work, we explore the behavior of the tilts at infinity for these cases, and find that their posteriors differ significantly from the posteriors at $f_{\rm ref}$ for some cases.
The second involves binaries whose spin configuration at the reference frequency is the endpoint of the up-down instability calculated by~\cite{Mould:2020cgc} (see also~\cite{Gerosa:2015hba, Lousto:2016nlp, Varma:2020bon}  
for studies of the up-down instability in general). We find that we are indeed able to recover the expected up-down tilt configurations at infinite separation, in agreement with the very recent study in~\cite{DeRenzis:2023lwa}.

The rest of the paper is organized as follows: In Sec.~\ref{sec:tilts-opt}, we describe the optimization scheme that determines the BBH spin configurations that show the largest individual tilt differences at $f_{\rm ref}$ and infinity. Using results from this, we outline our simulated observations in Sec.~\ref{sec:inj-study}, and present the results of our parameter estimation in Sec.~\ref{sec:results}. We conclude in Sec.~\ref{sec:conclusion} and give ancillary results in two appendices: Appendix~\ref{sec:appdxA} discusses the extreme sensitivity to input parameters we find in some cases and gives the parameters to the accuracy required to reproduce those results, while Appendix~\ref{sec:appdxB} gives tables with detailed parameter estimation results.

\section{Spin Tilts with largest differences at $f_{\rm ref}$ and infinity}
\label{sec:tilts-opt}

Ref.~\cite{Johnson-McDaniel:2021rvv} found very small deviation in tilt posteriors at infinity for GWTC-3 events with evidence for precession, the distributions were indistinguishable within statistical uncertainties at O3 detector sensitivities. 
As the detectors' sensitivities improve in the plus-era, we may start seeing significant differences in the posteriors at the reference frequency and at infinite separation. The goal here is to characterize for which binaries these deviations are most significant. 
In other words, for a BBH with a given value of total mass ($M$), mass ratio ($q=m_2/m_1$, where $m_1 \geq m_2$ are the individual masses), and spin magnitudes $(\chi_{1}, \chi_{2})$, what spin tilt configuration (characterized by $\theta_{1}, \theta_{2}, \phi_{12}$) shows the maximum change from $f_{\rm ref}$ to infinity? Further,
what is the dependence of the difference between the tilts at the reference frequency and at infinite separation on the binary's masses and spin magnitudes?

We measure the difference between the spin tilts at $f_{\rm ref}$ and infinity using the following quantity, $\delta$, which denotes the root mean square deviation in the cosines of primary and secondary spin tilt:

\begin{equation}
\delta := \sqrt{\left(\cos \theta_{1}^{\infty} - \cos \theta_{1}^{\rm ref}\right)^{2} + \left(\cos \theta_{2}^{\infty} - \cos \theta_{2}^{\rm ref}\right)^{2}}.
\end{equation} 

To start with, we checked how the spin angles changed for some individual cases. By choosing $1000$ binaries with random masses, spins, and tilt angles  
and then for each binary evolving the tilts to infinity while varying $\phi_{12}$, we found that the value of $\delta$ is always maximized when $\phi_{12}$ at $f_{\rm ref}$ is either close to $0$ or to $\pi$ (denoting the in-plane spin vectors pointing in the same or opposite directions). However, we observe a slight preference for $\phi_{12} \simeq \pi$ in binaries for which $\theta_{1}^{\rm ref} > \theta_{2}^{\rm ref}$ gives the maximum of $\delta$, and for $\phi_{12} \simeq 0$ otherwise, i.e., if $\theta_{2}^{\rm ref} > \theta_{1}^{\rm ref}$. If we solely use the precession-averaged evolution in computing the tilts at infinity, this maximization occurs when $\phi_{12}$ is \textit{exactly} either $0$ or $\pi$. With the hybrid evolution, $\delta$ is maximized at values slightly greater than $0$ or slightly less than $\pi$ (differences in $\delta$ values are at most $\sim 0.3$), especially for high total mass binaries with large spin magnitudes---making this presumably an effect of higher post-Newtonian (PN) terms used in the orbit-averaged part of the hybrid evolution. For cases where $\delta$ is maximized close to $\pi$, we compared the maximum function value with that at $\phi_{12} = \pi$, and found that the difference is less than $1\%$ for binaries with spin magnitudes $\chi_{1,2}$ of at least $0.5$ that we are interested in.
We thus restrict to considering either $\phi_{12} = 0$ or $\pi$, reducing the dimensionality of the search for optimized spin configurations to only the two tilt angles.

Since the hybrid evolution can take as much as $\sim 30$~s for close to equal-mass binaries~\cite{Johnson-McDaniel:2021rvv}, we use the \textsc{scipy}~\cite{2020SciPy-NMeth} implementation of the Nelder-Mead optimization algorithm~\cite{Nelder:1965zz} to maximize $\delta$ quickly and efficiently across a wide range of intrinsic binary parameters. To avoid settling on a local minimum, for each binary configuration, we run this algorithm using five random starting points, and take the $(\theta_{1}, \theta_{2})$ combination that attains the maximum $\delta$ value among these 
as the optimized binary tilt configuration. The optimization is performed with $\phi_{12}$ fixed to either $0$ or $\pi$, and we choose the $\phi_{12}$ value that gives the larger $\delta$.  In summary, our optimization code returns the spin angles $\{\theta_1^{\rm ref}, \theta_2^{\rm ref}, \phi_{12}^{\rm ref}\}$ that give the maximum difference in spin tilts between a given reference frequency and infinity (as measured using $\delta$), for a given set of $\{M, q, \chi_{1}, \chi_{2}\}$.

With this optimization code, we studied what binary configuration at $f_{\rm ref}$ leads to the largest $\delta$ value 
(also referred to as the ``peak tilt configuration" for a given binary), which in turn helped us choose the spin angles for our simulated BBH injections. We started with a wide-ranging set of BBHs having a fixed total mass of $M = 60 M_{\odot}$ and $f_\text{ref} = 20$~Hz, but mass ratio $q$ varying in the range of $[0.1,0.9]$ in increments of $0.1$, as well as an additional case of $q=0.99$, which takes us close to the $q=1$ (equal mass) limit where the tilts at infinity are not well defined (as discussed in, e.g.,~\cite{Johnson-McDaniel:2021rvv}). We took the component spin magnitudes $(\chi_{1}, \chi_{2})$ varying similarly but independently between $[0.1,0.9]$ with an additional value of $0.99$ corresponding to an almost maximally spinning black hole. We computed spin tilt configurations with the largest $\delta$ for these cases. 
In addition to getting spin tilt configurations, we also checked what kind of systems show particularly large $\delta$ values. 

We found that the largest tilt differences are obtained for binaries which are close to the up-down configuration at $f_{\rm ref}$, where the primary (secondary) tilt is aligned (anti-aligned) with the binary's orbital angular momentum $\vec{L}$. Among all BBHs, the largest $\delta$ values are
obtained for BBHs with close-to-equal masses ($q \sim 1$), and among these, systems with larger spin magnitudes give larger $\delta$. 
The maximum value of $\delta$ ($2.39$) was obtained for a binary with almost equal masses ($q=0.99$) and almost maximally equal spins ($\chi_1=\chi_2=0.99$), with the spin tilts at $f_{\rm ref}$ being $\cos{\theta_{1}^{\rm ref}}=0.99$ and $\cos{\theta_{2}^{\rm ref}}=-1.0$, which at infinity precess to $\cos{\theta_{1}^{\infty}}=-0.69$ and $\cos{\theta_{2}^{\infty}}=0.78$.\footnote{Assuming the effective spin is exactly conserved, $\delta$ is at most $2\sqrt{1 + q^2}$, which approaches its maximum value of $2\sqrt{2} \simeq 2.83$ as $q \to 1$.} 
On the other hand, binaries with unequal masses ($q\leq 0.5$) have $\delta$ values less than $0.61$ regardless of spin magnitudes. These binaries also have larger in-plane components of their optimized tilts, with $\cos{\theta_{1}^{\rm ref}}\in[-0.06, 0.48]$ and $\cos{\theta_{2}^{\rm ref}}\in[-0.2, 0.22]$.
The minimum value of $\delta$ ($0.027$) was obtained for the binary with the most lopsided masses ($q=0.1$), low and equal spins ($\chi_{1}=\chi_{2}=0.1$), and in-plane spins at $f_{\rm ref}$. In summary, higher $\delta$ values are obtained for binaries with higher $q$ (more equal masses) regardless of their spin magnitudes, while for a given mass ratio, $\delta$ is maximized for BBHs that have higher spins.

\section{Simulated GW signals and parameter estimation setup}
\label{sec:inj-study}

Having explored differences in spin tilts optimized for various individual BBHs, we use the results of the optimization analysis in the previous section to study the effect of spin evolution on the entire posterior distributions of such binaries in the plus era of detectors. We select a set 
of BBH systems to simulate 
in the Advanced LIGO and Advanced Virgo detectors with A+ and AdV+ sensitivity having the following properties: three different (redshifted) total mass values, $M \in \{50, 100, 200\} M_{\odot}$, three mass ratios, $q \in \{1/1.1, 1/3, 1/8\}$, and two dimensionless spin magnitudes, $\chi_{1}, \chi_{2} \in \{0.5,0.95\}$. The component black holes in a given binary could have either equal spins or unequal spins, giving a total of 36 simulated events. 
We pick moderate to large values for the component spins to increase the difference between the tilts at the reference frequency and infinity.

For each of these 36 binaries, values of $\{\theta_{1}, \theta_{2}, \phi_{12}\}$ at $f_{\rm ref}$ are obtained using our optimization scheme as described in Sec.~\ref{sec:tilts-opt}. We find that having the in-plane spins pointing in the same direction (i.e., $\phi_{12}=0$) at the reference frequency give the largest deviation in the tilts (as measured by $\delta$) in all cases. However, the maximum $\delta$ values for $\phi_{12}=\pi$ are not very different than those obtained for $\phi_{12}=0$ in most cases except for  $q = 1/1.1$ binaries. For binaries with more unequal mass ratios, the difference in $\delta$ does not exceed $\sim 6\%$, and lies within $2\%$ for most cases. For the $q=1/1.1$ cases, the minimum and maximum differences in $\delta$ are $0.04$ ($2.4\%$) and $0.69$ ($47\%$), respectively.

We observed that for $\phi_{12}=0$, the tilts at infinity have a very sensitive dependence on the exact tilt values at $f_{\rm ref}$ when using the hybrid evolution code---even small changes in the tilts at $f_{\rm ref}$ (or other binary parameters) give significant differences in the tilts at infinity. This sensitivity is seen most prominently for the comparable-mass binaries with smaller spins that give tilts at infinity close to the unstable up-down configuration---see Appendix~\ref{sec:appdxA} for more details. In those cases, using the tilts output by the optimization to their full $16$ decimal place precision gives tilts at infinity that are very close to the up-down spin configuration, but one obtains significantly different results 
when rounding off the inputs (e.g., we see a move towards in-plane tilts by $\sim 0.5$ even when rounding off to $6$ decimal places). For the other cases we consider, the change in the tilts at infinity is commensurate with the number of decimal places truncated in the input.

One also find that the results with the truncated tilt values are closer to those obtained when using the tilts with the full accuracy but using only the precession-averaged evolution, or the hybrid evolution with either a different PN approximant or only lower-order PN spin terms in the orbit-averaged evolution. In particular, one finds that using the $2.5$ and $2$PN spin terms in the orbit-averaged evolution gives hybrid evolution results for the tilts at infinity that are successively closer to the only precession-averaged evolution. However, if one redoes the optimization in the $\phi_{12}=0$ cases using a different PN approximant or only lower-order PN spin terms in the orbit-averaged evolution, one finds slightly different values of the tilts (differences of at most $\sim 0.03$) that give the same extreme sensitivity to the input parameters when using those settings for the orbit-averaged evolution. In all these cases, we find that the orbit-averaged evolution goes through the unstable up-down configuration, which explains the extreme sensitivity to the input parameters. (This instability is discussed further in Sec.~\ref{subsec:endpoint}.)

In summary, for each PN order or approximant used in the hybrid evolution, the peak tilt configuration when $\phi_{12}=0$ shows unstable behavior that is very sensitive to input parameters, due to passing through the unstable up-down configuration during the orbit-averaged evolution. (See Sec.~\ref{subsec:endpoint} for further discussion about the up-down instability, at the endpoints of which the spin vectors are collinear with $\phi_{12}=0$)
This sensitivity is not seen in tilt configurations where $\phi_{12}$ is set to $\pi$, though these configurations give $\sim 45\%$ smaller values of $\delta$ in the cases where there is hypersensitive behavior for $\phi_{12} = 0$, even though they give values of $\delta$ that differ by at most $\sim 6\%$ in other cases.

Owing to this hypersensitive behavior, we present results using both $\phi_{12}=0, \pi$ for comparable mass binaries. For unequal mass binaries, we use the optimized configuration at $\phi_{12}=0$ only when the resulting value $\delta$ obtained is higher by $2\%$ or more compared to the corresponding case with $\phi_{12}=\pi$. The final set of injected parameters is given in Appendix~\ref{sec:appdxB}.

The data for all binaries was simulated using the \textsc{IMRPhenomXPHM}~\cite{Pratten:2020ceb} waveform model (which includes precession and higher-order modes) with zero noise, and analyzed using the A+/AdV+ noise curves~\cite{KAGRA:2013rdx} (using the more sensitive AdV+ noise curve) to compute the likelihood. These \textit{injections} of binary signals into plus-era detectors were made at a fixed luminosity distance of $400 \text{ Mpc}$ (a choice motivated by the distance estimate of GW150914~\cite{LIGOScientific:2016aoc}, though it is on the lower side of the distances measured for GWTC-3 events) and at an inclination angle of $60^\circ$. Other extrinsic properties such as the sky locations, coalescence phase, and polarization angle of the signal were chosen randomly and kept fixed for all events as follows: \{right ascension: $5.285$, declination: $0.908$, polarization angle: $3.989$, coalescence phase: $2.658$\}, all in radians. The simulation geocentric merger GPS time was set at $1257995721.214$ seconds.

The reference frequency $f_{\rm ref}$ was chosen to be the same as the low-frequency cutoff, which was selected according to the total mass to be $15$~Hz for the 
$50 M_{\odot}$ case, and $10$~Hz for the $100 M_{\odot}$ and $200 M_{\odot}$ cases so as to keep a computationally feasible number of cycles in the signal while ensuring a negligible contribution to the signal-to-noise ratio (SNR) from frequencies below the cut-off.
Similarly, we considered the SNR contribution from higher-order modes while determining the sampling frequency for our simulated signals so as to keep the signal duration short and reduce the computational cost of the parameter estimation. The corresponding Nyquist frequencies for the $\{50, 100, 200\} M_\odot$ simulations were $\{1024, 512, 256\}$~Hz respectively. 
A roll-off factor of $0.875$ was applied to these before setting the high-frequency cutoffs to account for the effects of a window function (as discussed in Appendix~E of~\cite{LIGOScientific:2021djp}).
Given that the injections are made using plus-era noise curves~\cite{KAGRA:2013rdx} at the same luminosity distance, the SNR ranges for the $50M_\odot$, $100M_\odot$, and $200M_\odot$ cases are $[47, 89]$, $[54, 148]$, and $[87, 272]$, respectively. The variation in SNR comes primarily  from the differences in the mass ratio, and whether the spins are aligned or anti-aligned with the orbital angular momentum---quantities which dictate the number of cycles in the sensitive band.

All these injections were recovered using the \textsc{IMRPhenomXPHM} waveform model and the standard publicly-available \textsc{parallel bilby}~\cite{pbilby_paper, bilby_paper} inference code with the \textsc{dynesty} nested sampler~\cite{dynesty_paper}, with the `nact' parameter~\cite{Romero-Shaw:2020owr} set to $50$. 
Following the LVK parameter estimation analyses~\cite{LIGOScientific:2021djp}, we choose priors that are uniform in redshifted component masses and isotropic spin orientations. We assume a uniform distribution of sources in time and comoving volume (with the exception that we use the Euclidean distance prior as opposed to a cosmological one in~\cite{LIGOScientific:2021djp}).

The parameter estimation analysis gives us the posterior distributions of binary's masses and spin parameters at $f_{\rm ref}$. We compute the posterior distribution of tilts at infinity by evolving these distributions to infinite separation using our hybrid orbit + precession-averaged evolution code~\cite{Johnson-McDaniel:2021rvv}. In the next section, we examine the differences in tilt posteriors at $f_{\rm ref}$ and infinity and characterize the BBH parameters that show the maximum differences.

In addition to the above mentioned injections, we also consider the parameter estimation results from the study of precessional morphologies in~\cite{Johnson-McDaniel:2023oea}, which use both \textsc{IMRPhenomXPHM} and the numerical relativity surrogate model \textsc{NRSur7dq4}~\cite{Varma:2019csw}. These injections involve two total masses $M = \{20, 75\} M_{\odot}$ with a fixed mass ratio of $q = 1/1.2$, and three equal spin magnitudes $\chi_1 = \chi_2 = \{0.25, 0.75, 0.95\}$ with a fixed SNR of $89$ and an inclination angle of $60^\circ$ in the same plus-era LIGO-Virgo network we consider in this paper. 
While~\cite{Johnson-McDaniel:2023oea} also considers some SNR $22$ cases, we do not consider these here. Additionally, we analyze a set of four injections using the up-down instability endpoint tilt angles given in~\cite{Mould:2020cgc}, which consist of binaries having total mass $M \in \{20,50,100,200\}M_\odot$ with close-to-equal masses $(q=1/1.1)$ and high spins $(\chi_{1} = \chi_{2} = 0.95)$.

\section{Results}
\label{sec:results}

The high SNR signals in plus-era detectors give much tighter constraints on the individual tilts of black holes in our injections set than for any of the events in GWTC-3, as was already noted in~\cite{Knee:2021noc}. For brevity, we present the injected binary parameters as well as the medians and 90\% credible intervals (CIs) of the recovered parameters in Appendix~\ref{sec:appdxB}. Here we discuss some specific observations we made from our parameter estimation analysis. 
The most well-constrained spin tilts we obtain are for the binary having \{$M=200 M_{\odot}, q=1/1.1, \chi_{1}=\chi_{2}=0.95$\}, 
with the 90\% CI for $\theta_{1}$ and $\theta_{2}$ having widths of $1.5^\circ$ and $1.6^\circ$, respectively. On average, the widths of the 90\% CI of the recovered spin tilts in our simulations are $15.8^\circ$ ($\theta_{1}$) and $42.8^\circ$ ($\theta_{2}$), which are significantly better than GWTC-3 events. For example, GW200129\_065458, an event with median SNR of 27 that shows some evidence for precession~\cite{LIGOScientific:2021djp} has 90\% spin tilt CIs of $76.14^\circ$ ($\theta_{1}$) and $118.15^\circ$ ($\theta_{2}$). 

\begin{figure*}[h]
\centering
\subfloat{
\includegraphics[width=0.98\linewidth]{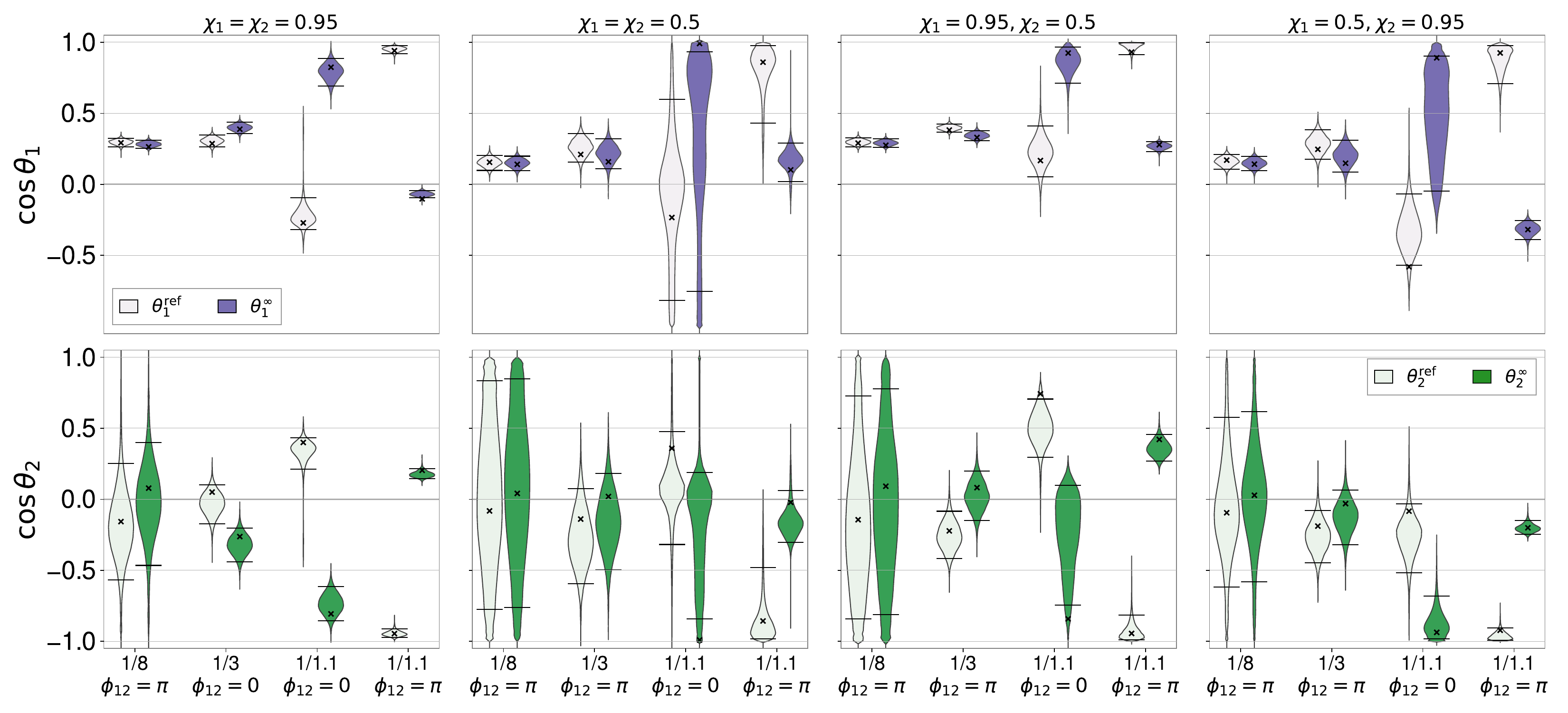}
}
\caption{\label{fig:M50} Violin plot showing the spin tilt distributions at $f_{\rm ref}$ and infinity for various binary configurations having a total mass of $50 M_{\odot}$. The horizontal lines on each violin represent bounds of the $90\%$ CI of the distribution, while the crosses represent the injected tilt values at either $f_{\rm ref}$ or infinity. Note that for injections close to the aligned and anti-aligned configuration, the posteriors rail against the prior boundary of $\cos{\theta} = \pm 1$. 
}
\end{figure*}

\begin{figure*}[h]
\centering
\subfloat{
\includegraphics[width=0.98\linewidth]{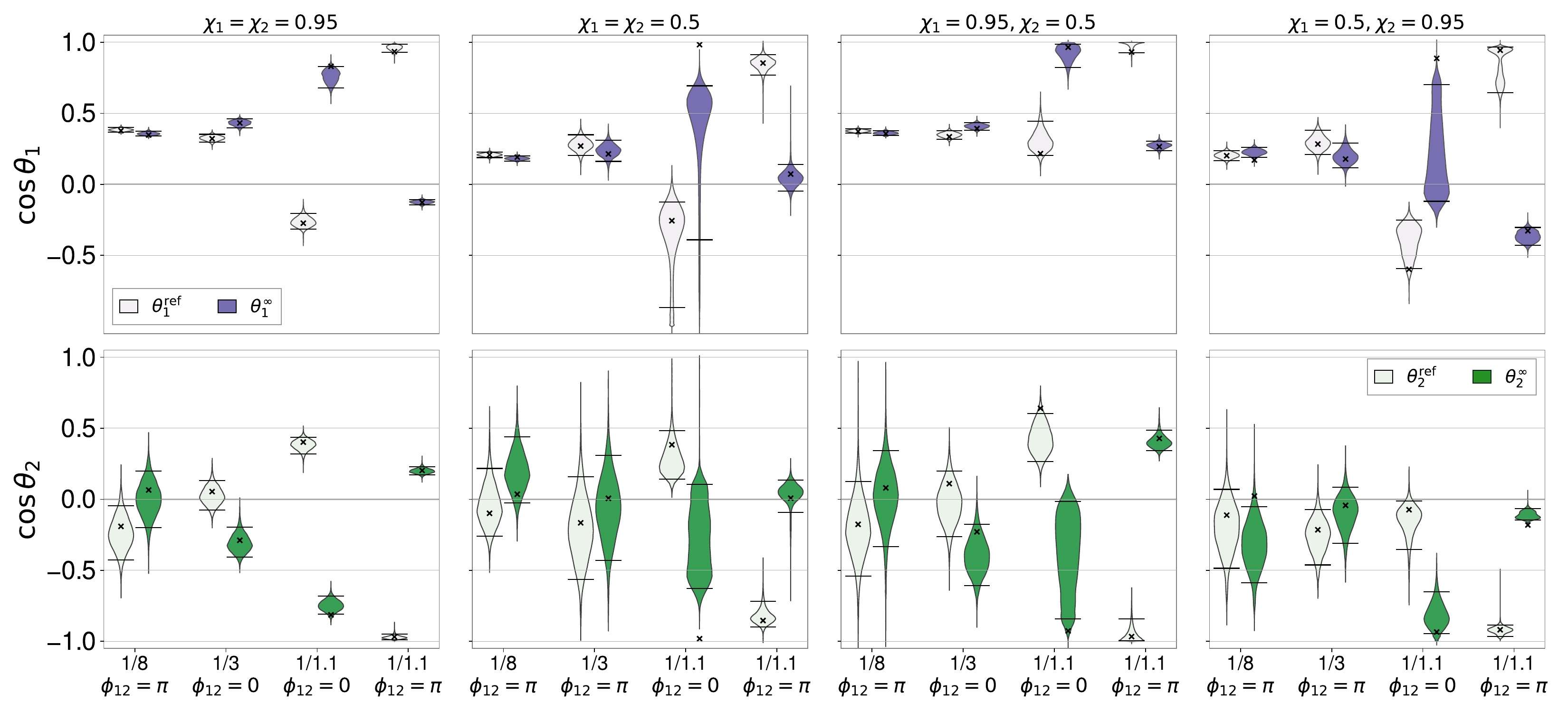}
}
\caption{\label{fig:M100} Violin plot showing the spin tilt distributions at $f_{\rm ref}$ and infinity for various binary configurations having a total mass of $100 M_{\odot}$. For more details, refer to the caption of Fig.~\ref{fig:M50}.
}
\end{figure*}

\begin{figure*}[h]
\centering
\subfloat{
\includegraphics[width=0.98\linewidth]{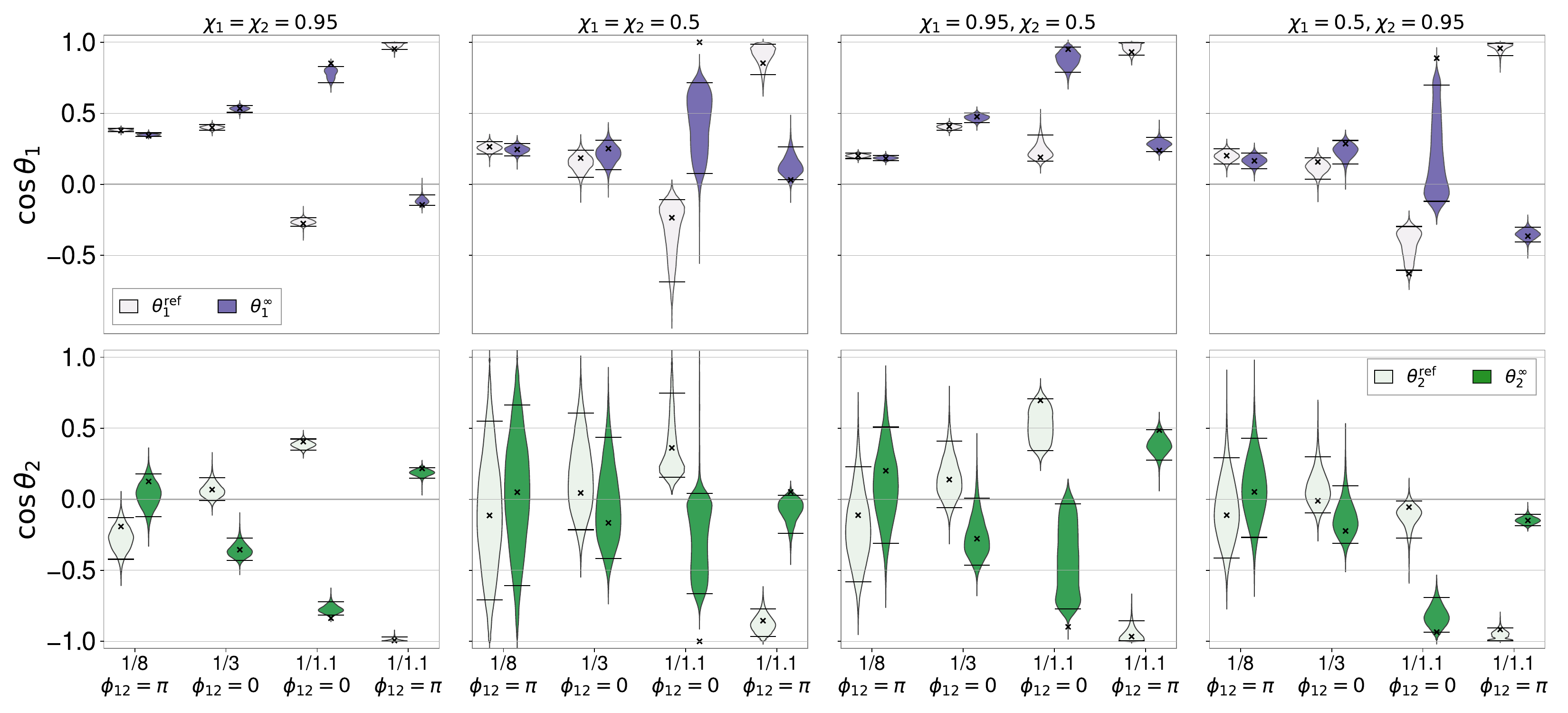}
}
\caption{\label{fig:M200} Violin plot showing the spin tilt distributions at $f_{\rm ref}$ and infinity for various binary configurations having a total mass of $200 M_{\odot}$. For more details, refer to the caption of Fig.~\ref{fig:M50}.
}
\end{figure*}

With these tighter constraints, we observe that for some binaries, their tilt posteriors show significant deviations when evolved to infinity. The differences in tilt posteriors at $f_{\rm ref}$ and infinity for our injection set are shown as violin plots in Figs.~\ref{fig:M50}--\ref{fig:M200}.

\begin{figure*}[h]
\centering
\subfloat{
\includegraphics[width=0.98\linewidth]{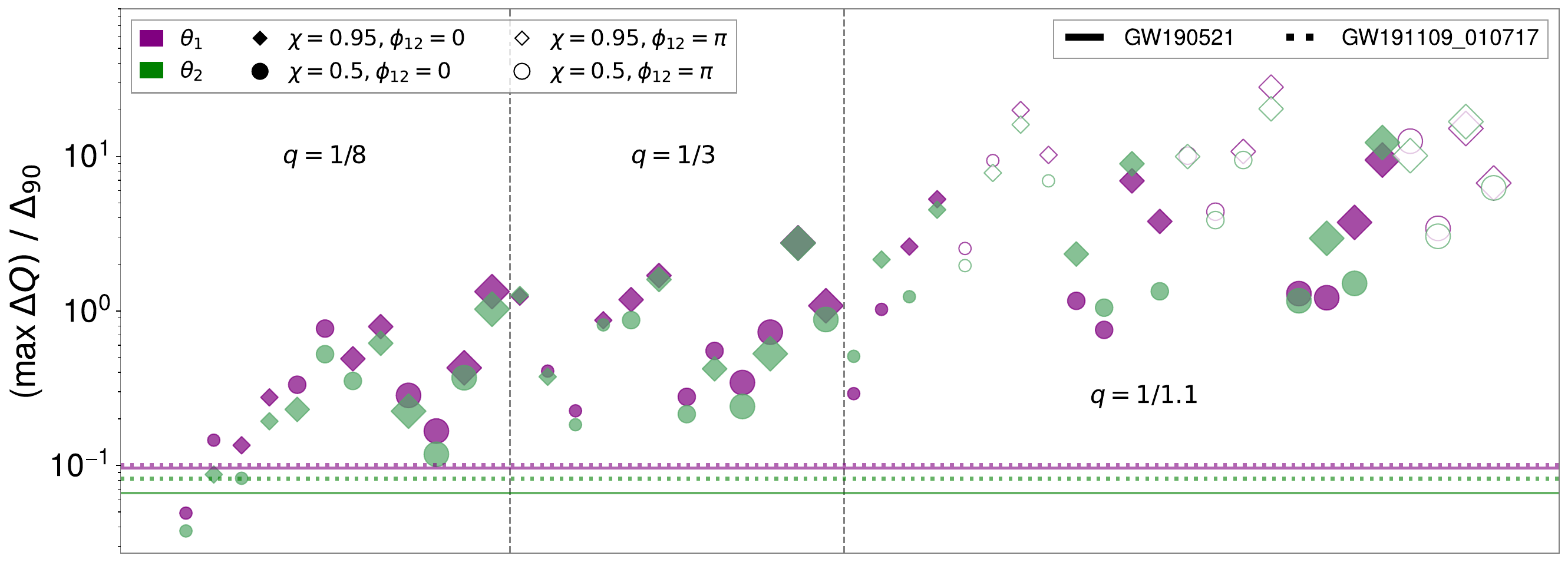}
}
\caption{\label{fig:maxDQ} The scaling of tilt differences in terms of $(\max \Delta Q)/\Delta_{90}$ for simulated binaries that are grouped by mass ratio, $q$ increasing from unequal masses to the left, towards comparable masses to the right. Each binary is denoted by two markers which represent individual BH spin magnitudes ($\chi$): circles for moderate spins and diamonds for high spins. Unfilled markers represent $q = 1/1.1$ binaries with $\phi_{12}=\pi$. Within each $q$-panel, the binaries are ordered by increasing total mass, $M$, with heavier binaries towards the right also indicated by larger marker size. For a given $q$ and $M$, the binaries are ordered by their primary spin magnitude. The dashed (solid) horizontal lines indicate $(\max \Delta Q)/\Delta_{90}$ for GW191109\_010717 (GW190521), two GWTC-3 events representative of the largest deviations in tilt posteriors at O3 sensitivities. 
}
\end{figure*}

The degree to which the tilt distributions differ can be quantified by the max $\Delta Q$ parameter used in~\cite{Johnson-McDaniel:2021rvv}. If two distributions labelled I and II are to be compared, max $\Delta Q$ is defined by
\begin{equation}
\label{eqn:maxDQ}
\max \Delta Q := \max \left(|Q_{5}^{\rm I}-Q_{5}^{\rm II}|, |Q_{50}^{\rm I}-Q_{50}^{\rm II}|, |Q_{95}^{\rm I}-Q_{95}^{\rm II}| \right),
\end{equation}
where $Q_{n}^{Y}$ is the $n\%$ quantile of the $Y$th distribution. Thus, $\max \Delta Q$ is the maximum absolute shift among the $5$\%, $50$\%, and $95$\% quantiles of the two distributions (i.e., the differences in the median and the $90$\% credible interval around it). To account for the different detector sensitivities and consequently better constraints on tilts, we modify this quantity by dividing it by the width of the $90\%$ CI at infinity ($\Delta_{90}$) of each posterior distribution. 
The largest $(\max \Delta Q) / \Delta_{90}$ values seen in the GWTC-3 catalog so far are $0.1$ ($0.08$) for $\cos \theta_{1}$ ($\cos \theta_{2}$) in the event GW191109\_010717. For GW190521, the primary (secondary) tilt $(\max \Delta Q) / \Delta_{90}$ values are $0.09$ ($0.06$). For comparison, Fig.~\ref{fig:maxDQ} shows the $(\max \Delta Q) / \Delta_{90}$ values for our injections ordered by their mass ratio. 

We summarize the results from these plots (Figs.~\ref{fig:M200}--\ref{fig:maxDQ}) by considering the effects of the various binary parameters that impact these spin tilt posterior differences:

\textit{Effect of mass ratio:} We observe that the binary mass ratio plays the dominant role in determining the deviation in tilt posteriors. The largest differences, where the 90\% CIs are  disjoint, are seen in close-to-equal-mass ($q = 1/1.1$) cases with $\phi_{12}=\pi$. These binaries also have the highest $(\max \Delta Q)/\Delta_{90}$ values. Binaries having $\phi_{12}=0$ and comparable masses, on the other hand, show smaller deviations in their tilt posteriors, in spite of the individual tilts at infinity corresponding to their injected tilt values giving larger $\delta$ values than their $\phi_{12}=\pi$ counterparts. This is because of two factors: First, we see that the spin tilts for binaries with $\phi_{12}=\pi$ are more tightly constrained due to greater spin-induced modulations.
Secondly, the issue of sensitivity to the initial tilt values outlined in Sec.~\ref{sec:inj-study} leads to individual spin tilt posterior samples evolving to considerably different tilts at infinity compared to those given by the precise injected optimized tilt configuration, even if the samples' parameters are close to the injected values. For example, this is seen in the $\{M=200 M_{\odot}$, $q=1/1.1$, $\chi_1=\chi_2=0.5\}$ binary with $\phi_{12}=0$ in Fig.~\ref{fig:M200}, where the injected tilts at infinity are entirely outside the evolved posterior distributions. As discussed in Appendix~\ref{sec:appdxA}, one obtains tilts at infinity that are closer to the posteriors when using fewer decimal places for the injected parameters, though still lying in the tails of the distributions. 
As a result, the deviation seen in tilt distributions as well as the $(\max \Delta Q)/\Delta_{90}$ values for these cases are slightly lower than comparable mass binaries with $\phi_{12}=\pi$. 
These differences are still larger than binaries with the intermediate mass ratio ($q = 1/3$), while more unequal mass ($q = 1/8$) binaries have significant overlaps in their posteriors at $f_{\rm ref}$ and infinity. The overlap of posteriors in the  $q = 1/8$ case is small (with disjoint $90\%$ CIs) only for the $200M_\odot$ binary with $\chi_1 = \chi_2 = 0.95$. 

\textit{Effect of spins:} For binaries with a given mass ratio, the largest difference in tilt posteriors is seen for binaries with high primary spins, either in the equal spin ($\chi_{1} = \chi_{2} = 0.95$) or unequal spin cases ($\chi_{1} = 0.95$, $\chi_{2} = 0.5$). 
The recovery of spin tilts for the $q=1/1.1$ binaries is poorer when $\phi_{12}$ is set to $0$ as opposed to $\pi$. We also note that for some of these $\phi_{12}=0$ cases, the injected tilt values lie outside the $90\%$ CI of their recovered posteriors because of one or both individual in-plane spin components 
of either BH not being well-constrained at $f_{\rm ref}$ in the parameter estimation. 
Further, we observe that cases wherein the spin magnitude (especially for the primary) is low have their tilt posteriors at infinity deviating significantly from the expected tilts at infinity computed using the hybrid code, due to the same sensitive behavior outlined in Sec.~\ref{sec:inj-study}.

\textit{Effect of total mass:} As evident from the violin plots, all trends seen in the mass ratio are consistent across injections with varying total masses, while the spins show slight differences as outlined above. Binaries with higher masses give louder signals and hence have better-constrained tilts, which leads to differences in their posterior distributions at infinity being more significant [as seen in the $(\max \Delta Q) / \Delta_{90}$ trend in Fig.~\ref{fig:maxDQ}]. Spin tilt posteriors of high-mass binaries also show differences when evolved using the precession-averaged and hybrid spin evolution approaches, as we describe in detail in Sec.~\ref{subsec:hyb_vs_prec}.

\begin{figure}[h]
\centering
\subfloat{
\includegraphics[width=0.95\linewidth]{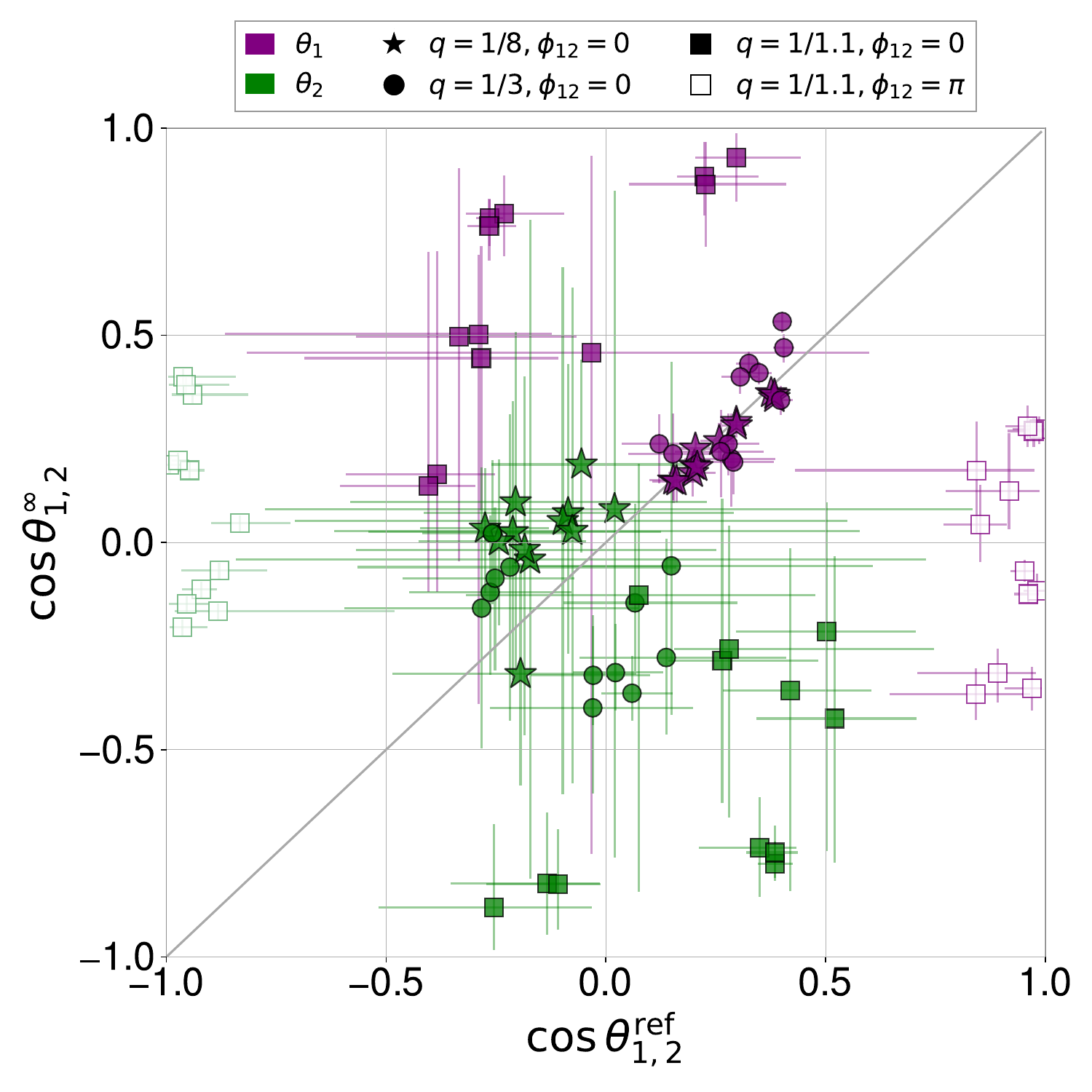}
}
\caption{\label{fig:cos_tilts_scatter} Comparison of the medians of the cosines of the spin tilts of the binaries at the reference frequency (horizontal axis) with those at infinity (vertical axis). The error bars denote the $90\%$ CIs. Unfilled markers represent $q = 1/1.1$ binaries with $\phi_{12}=\pi$. BBHs lying along the diagonal ($\cos\theta^\text{ref} = \cos\theta^\infty$) line have very similar posterior tilt distributions at $f_{\rm ref}$ and infinity. Cases where the error bars cross the diagonal line are ones where the $90\%$ CIs of the two distributions have significantly overlap. 
}
\end{figure}

Figure \ref{fig:cos_tilts_scatter} summarizes our results for all injections by comparing the medians of the cosines of the spin tilts at $f_{\rm ref}$ and infinity. The error bars represent the 90\% CI of the posterior distributions. A majority of points either lie very close to the $\cos \theta_{1,2}^\text{ref} = \cos \theta_{1,2}^\infty$ line, or have it within their 90\% CIs. Notably, all outliers away from this line have smaller error bars and represent close-to-equal-mass binaries. Among these, binaries with $\phi_{12}=\pi$ at $f_{\rm ref}$ (unfilled square markers) have their primary spins close to aligned with $\vec{L}$, while their secondary spins 
are close to anti-aligned. At infinity, both spins are instead in a more in-plane configuration, being scattered around $\cos{\theta_{1,2}^{\infty}} \simeq 0$. In contrast, the spins for the $\phi_{12}=0$ binaries (filled square markers) move from closer to in-plane at $f_{\rm ref}$ towards the aligned-antialigned configuration at infinity. Binaries with unequal masses ($q=1/8$) have poor constraints on their secondary spin tilts (e.g., the $90\%$ C.I. for the $M=50 M_{\odot}$ binary with equal spins of $\chi=0.5$ spans $107.4^\circ$), and hence have larger error bars. %

\begin{figure}[h]
\centering
\subfloat{
\includegraphics[width=0.95\linewidth]{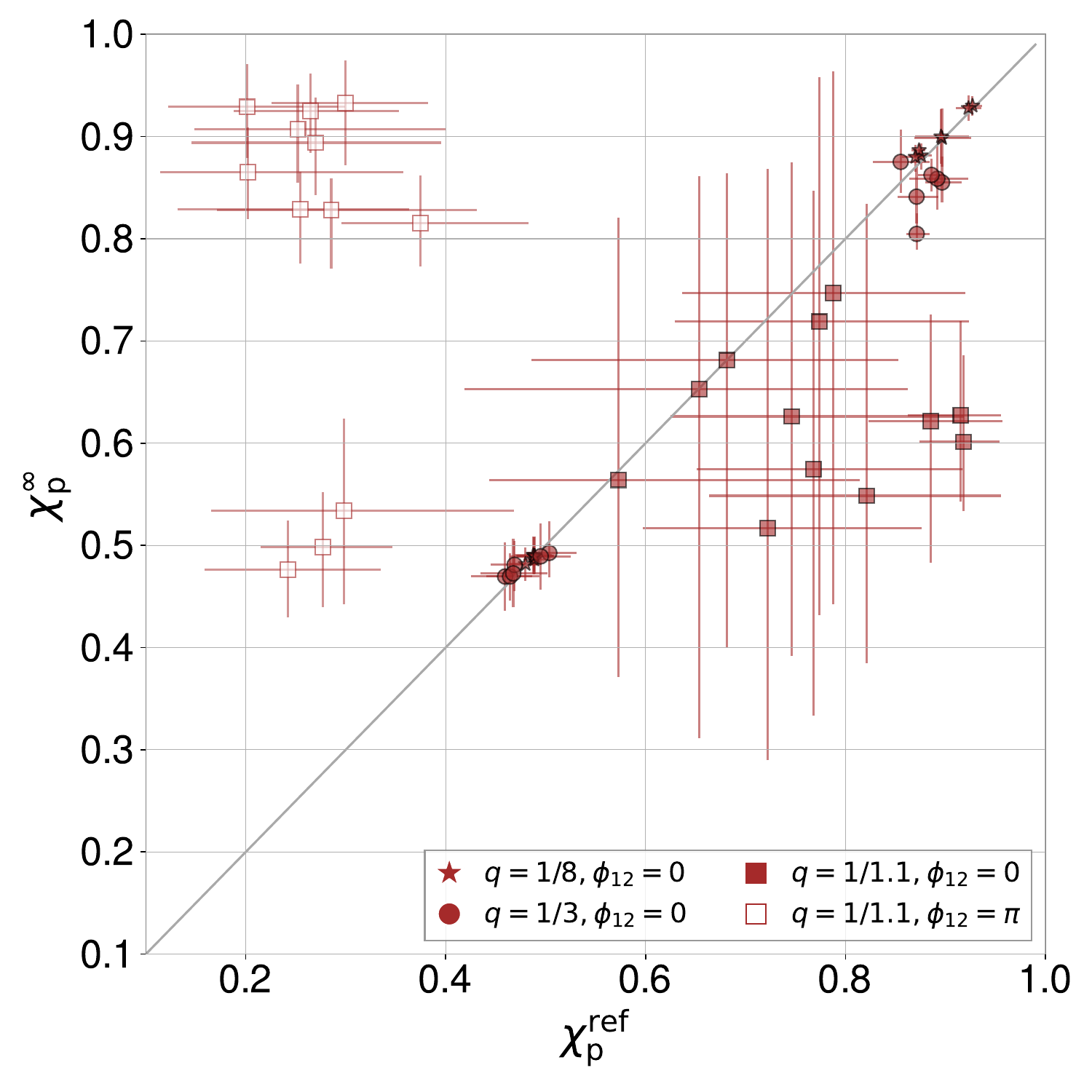}
}
\caption{\label{fig:chi_p_scatter} Comparison of the effective precession spin parameter ($\chi_{\rm p}$) posterior distributions at $f_{\rm ref}$ (horizontal axis) and at infinity (vertical axis). The error bars denote the $90\%$ CIs. Marker styles denote binaries with different mass ratios. Unfilled markers represent $q = 1/1.1$ binaries with $\phi_{12}=\pi$. Cases where the error bars cross the diagonal line are ones where the $90\%$ CIs of the two distributions have significantly overlap. 
}
\end{figure}

Finally, we studied the posterior distributions of the effective spin $\chi_{\rm eff}$~\cite{Santamaria:2010yb, Racine:2008qv} and effective precession spin parameter $\chi_{\rm p}$~\cite{Hannam:2013oca, Schmidt:2014iyl} for our binaries when evolved to infinite separation. $\chi_{\rm eff}$ is a conserved quantity in the 2PN precession-averaged spin evolution equations, and is approximately conserved in the hybrid evolution that uses higher-PN terms in the orbit-averaged evolution (see~\cite{Johnson-McDaniel:2021rvv}). As expected, we do not see any significant deviations in the $\chi_{\rm eff}$ posteriors for our injections. When it comes to $\chi_{\rm p}$, however, the picture is different. A majority of our $q=1/1.1$ binaries show significant precession, and among these, there are two distinct populations corresponding to $\phi_{12}=0$ or $\pi$. The $\phi_{12}=\pi$ binaries, denoted by unfilled square markers in Fig.~\ref{fig:chi_p_scatter}, start from an anti-aligned configuration at $f_{\rm ref}$ with small $\chi_p$ and transition towards more in-plane spins at infinity giving large $\chi_p$ values there. The $\phi_{12}=0$ binaries on the other hand start with slightly higher $\chi_{\rm p}$ values at $f_{\rm ref}$ than at infinite separation. The difference in $\chi_{\rm p}$ is again not significant for more unequal mass binaries.

\subsection{Different Spin Morphologies}
\label{subsec:morph}

Precessing BBHs can be classified into three different morphologies according to the patterns traced by the black hole spins over a precessional cycle. These are generalizations of BBH spin-orbit resonances~\cite{Schnittman:2004vq}, previously studied through their effects on GW evolution~\cite{Gupta:2013mea, Gerosa:2014kta, Trifiro:2015zda, Afle:2018slw, Varma:2021xbh} (see also work for eccentric binaries in~\cite{Phukon:2019gfh}), that are characterized using the effective potential formalism described in~\cite{Gerosa:2015tea, Kesden:2014sla}. 
Spin morphologies can be characterized using the evolution of $\phi_{12}$ as: (i) The L$0$ morphology where $\phi_{12}$ librates around $0$ (the two in-plane spin components point roughly towards the same direction), (ii) The L$\pi$ morphology where $\phi_{12}$ librates around $\pi$ (the two in-plane spin components point roughly in opposite directions), and (iii) the C morphology where $\phi_{12}$ circulates over the whole range $[-\pi, \pi]$. While all systems are in the C morphology at infinite separation, some transition into L$0$ or L$\pi$ closer towards merger.

In \cite{Johnson-McDaniel:2023oea}, the authors showed that it will be possible to infer the true morphology at the reference frequency with high statistical confidence for a sufficiently loud GW event using Bayesian model selection. This inference is stronger if the system did not recently undergo a transition between two morphologies at the reference frequency (equivalent to being far away from any of the boundaries between morphologies in the binary's parameter space). Here, we take posterior samples from that study and evolve them backward in time to see how the inferred tilt distributions corresponding to a given morphology change when evolved to infinity. For each combination of injected mass and spins, the spin angles are chosen either to lie centrally in the region of parameter space corresponding to each of the C, L$0$, or L$\pi$ morphologies, or to lie close to (and on either side of) the C-L$0$ and C-L$\pi$ morphology boundaries. For binaries in the librating morphologies, the cases close to the boundary between morphologies correspond to the situations where the binary has been in the librating morphology for only a short time before it reaches the reference frequency. In all cases, these morphologies are computed at the reference frequency of $20$~Hz.

\begin{figure*}[h]
\centering
\subfloat{
\includegraphics[width=0.95\linewidth]{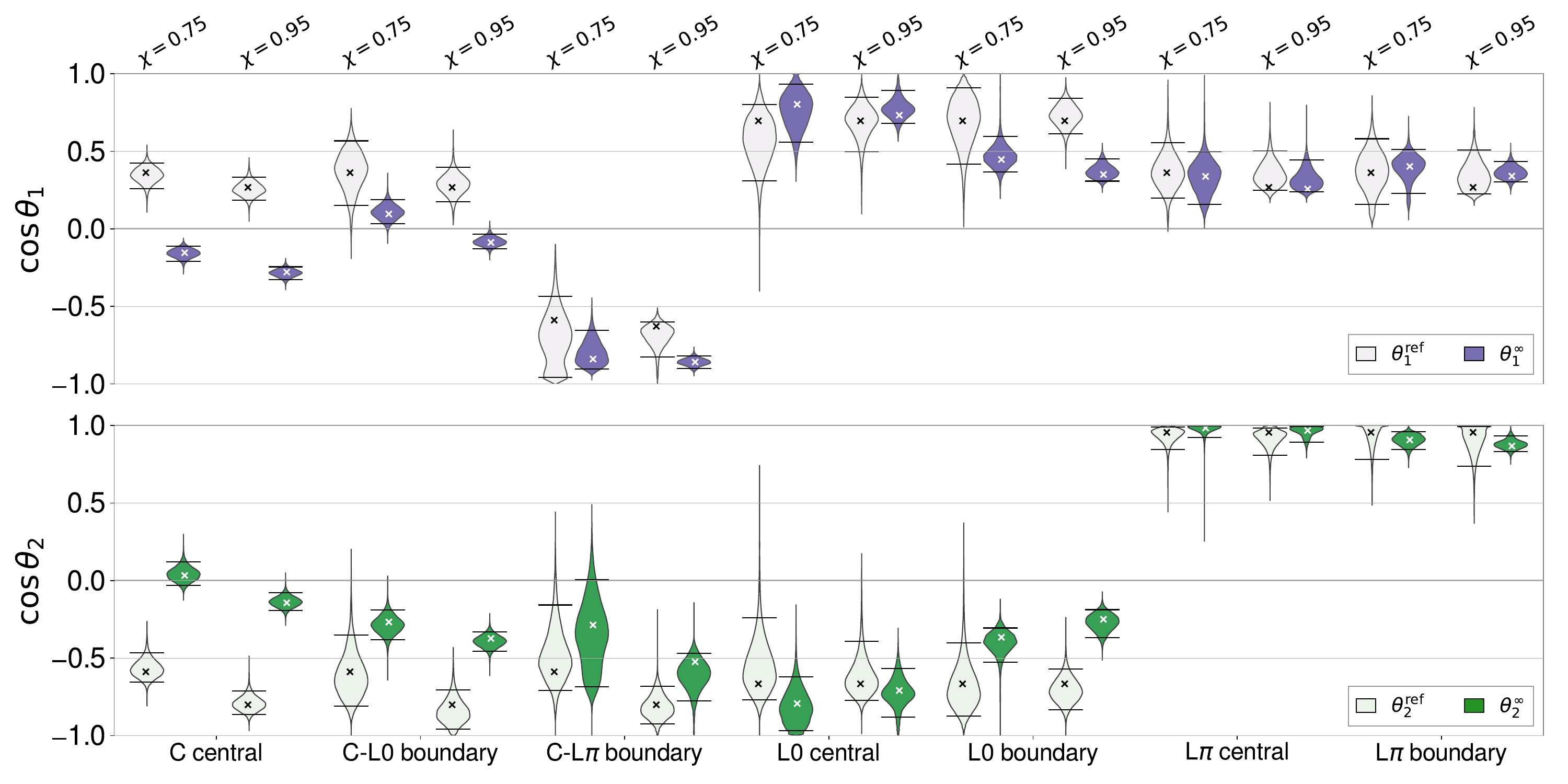}
}
\caption{\label{fig:morphology_violinplot_M20}
Comparison of spin tilts at $f_{\rm ref}$ and infinity for tilt posteriors belonging to different spin morphologies for total mass $M=20 M_{\odot}$. For each morphology, results for spin magnitudes $(\chi_1=\chi_2=\chi)$ $0.75$ and $0.95$ are presented side-by-side as indicated in the upper panel. 
}
\end{figure*}

\begin{figure*}[h]
\centering
\subfloat{
\includegraphics[width=0.95\linewidth]{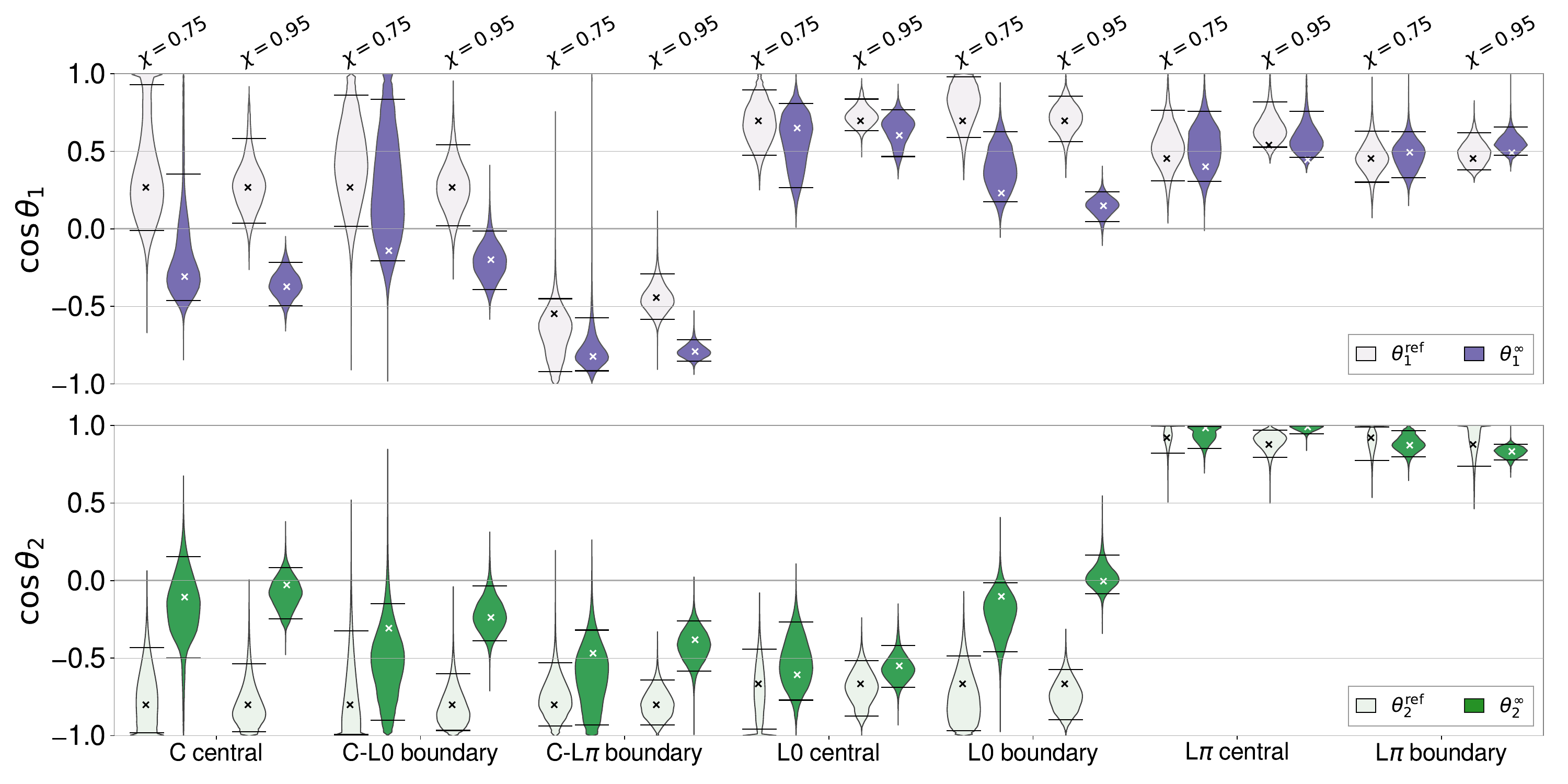}
}
\caption{\label{fig:morphology_violinplot_M75} 
Comparison of spin tilts at $f_{\rm ref}$ and infinity for tilt posteriors belonging to different spin morphologies for total mass $M=75 M_{\odot}$. For each morphology, results for spin magnitudes $(\chi_1=\chi_2=\chi)$ $0.75$ and $0.95$ are presented side-by-side as indicated in the upper panel. The particularly long violin tail for the primary tilt at infinity in the C-L$\pi$ boundary $\chi=0.75$ case is due to two outlying posterior samples with $\cos{\theta_{1}^\infty} \simeq 0.97$. 
}
\end{figure*}

The behavior of the tilt distributions at the reference frequency of $20$~Hz and at infinity for these injections is summarized as violin plots in Figs.~\ref{fig:morphology_violinplot_M20} and \ref{fig:morphology_violinplot_M75} for BBHs with total mass $20M_{\odot}$ and $75 M_{\odot}$ respectively. The results shown are for $\chi_1=\chi_2 = \chi=0.95, 0.75$. We do not present the results for the $\chi = 0.25$ cases here because for these cases the distributions at $f_{\rm ref}$ and infinity are broad and almost identical regardless of the morphology, with only a small deviation in the location of the peak seen in the C central cases.

Describing the low mass ($M=20M_\odot$) cases first, for both moderate ($\chi=0.75$) and high ($\chi=0.95$) spins, we observe the largest deviations in tilt posteriors for binaries in the C morphology at $f_{\rm ref}$, followed by the L$0$ and L$\pi$ morphologies. The differences are particularly prominent for injections within the central C region (where the distributions are completely disjoint) and close to the C-L$0$ boundary (where there are slight overlaps). Thus, one can obtain significant differences in the posteriors at infinity and the reference frequency even in cases that were not specifically designed to obtain large differences. Interestingly, for these cases the spins precess from an anti-aligned configuration at infinity towards an in-plane one at the reference frequency, the opposite of most of the cases that we considered that give the maximum differences. At the C-L$\pi$ boundary, the posteriors have a greater overlap.
The L$0$ morphology posteriors show slight deviations towards anti-alignment within the central part of the region, and larger but opposite deviations towards an in-plane configuration at the boundary.
For the L$\pi$ morphology, binaries both within the  region and near its boundary only show very small deviations from $f_{\rm ref}$ to infinity.

We find that the spin tilt distributions for the high mass ($M = 75 M_{\odot})$ injections show similar trends as above.  
In particular, we find notable differences in the posteriors for injections with both spin magnitudes in the C central cases as well as the C-L$0$ and L$0$ boundary cases. Binaries in the central part of the L$0$ morphology show a smaller shifts, from anti-alignment towards in-plane spins. 
Interestingly, in the C-L$\pi$ boundary $\chi = 0.95$ case, the medians of the $\cos{\theta_{1}}$ and $\cos{\theta_{2}}$ distributions switch places when evolved from $f_{\rm ref}$ to infinity. Overall, the L$\pi$ morphology shows only small shifts in the primary tilt distributions both within the central region and its boundary with C, while the differences are more pronounced in the better-constrained secondary tilts. All of the above trends are stronger in the high spin ($\chi=0.95$) cases as compared to the $\chi=0.75$ ones. 

The availability of \textsc{NRSur7dq4}~\cite{Varma:2019csw} numerical relativity surrogate evolution posterior samples for six of these injections allowed us to test for any systematic effect of the choice of $f_{\rm ref}$ on the hybrid evolution to infinity for plus-era posteriors, by using the surrogate evolution to evolve the binary backwards within the range of validity of the surrogate model. We already checked in~\cite{Johnson-McDaniel:2021rvv} that there is no significant effect from the surrogate evolution for the GW190521 posteriors. Starting from $f_{\rm ref}=20$ Hz, we evolved the posteriors for these six binaries (with total mass $75 M_{\odot}$ and spin magnitudes $\chi=\{0.25,0.75,0.95\}$) back to a time of $4200M$ before merger (corresponding to transition frequencies of $\sim 14$ Hz) using the surrogate evolution, before switching to the hybrid evolution to obtain the tilt posteriors at infinity. 

We observed no significant differences between these and posteriors computed using hybrid evolution from $20$ Hz, with only very minute shifts in the posteriors in the highly spinning cases. However, in the cases where we have \textsc{NRSur7dq4} results, there are also no significant differences when using only the precession-averaged evolution as opposed to the hybrid evolution, the posteriors of which can differ for certain binaries as outlined later in Sec.~\ref{subsec:hyb_vs_prec}. Thus, it is possible that we would find a larger difference when using the surrogate evolution for other binaries. Additionally, we see no systematic differences when comparing the \textsc{IMRPhenomXPHM} and \textsc{NRSur7dq4} posteriors at infinity for these six cases that cannot be attributed to a difference in the two waveform posteriors at $f_{\rm ref}$. The posterior distributions of the tilts at infinity inferred using the two waveform models generally agree well, with \textsc{NRSur7dq4} results generally being somewhat better constrained and showing small to moderate shifts in the positions of the peaks. The largest difference is in the C-L$0$ boundary case with $\chi=0.75$, where the tilts at infinity distributions inferred with \textsc{NRSur7dq4} are notably better constrained (particularly for the primary) and peak closer to the injected values than those obtained with \textsc{IMRPhenomXPHM}.

\subsection{Endpoints of the Up-Down Instability}
\label{subsec:endpoint}

As discussed in, e.g.,~\cite{Mould:2020cgc}, the PN spin evolution equations have four equilibrium, non-precessing solutions corresponding to both the primary and secondary spin vectors oriented exactly along the direction of $\vec{L}$. These configurations are: up-up, down-down, down-up, and up-down, where up (down) refers to the component spin being aligned (anti-aligned) with $\vec{L}$. Out of these, the up-up, down-down, and down-up configurations are stable equilibria, with small tilt perturbations remaining close to alignment over the course of spin evolution. The up-down configuration on the other hand exhibits an instability that is present for binary separations below a critical value. However, even after the onset of instability, the spin vectors attain well-defined endpoints such that they are oriented in the same direction (so $\phi_{12}^\text{ude} = 0$) and with tilt angles given by~\cite{Mould:2020cgc}
\begin{equation}
\label{eqn:ude}
\cos\theta_1^\text{ude} = \cos\theta_2^\text{ude} = \frac{\chi_{1} - q\chi_{2}}{\chi_{1} + q\chi_{2}}.
\end{equation}
The endpoints thus depend only on the mass ratio and spin magnitudes of the two component black holes. Here, we study the spin tilt evolution of BBHs starting from these endpoints to infinite separation.

Our injected tilts at $f_{\rm ref}$ were computed using Eq.~\eqref{eqn:ude}, and the injected value of $\phi_{12}$ was set to be $0$. The other binary parameters are given in Sec.~\ref{sec:inj-study}. 
We use a reference frequency of $20$ Hz, which is well above the frequency of the onset of the up-down instability [see Eq.~(21) in~\cite{Mould:2020cgc}, which gives the binary separation at onset of instability; the corresponding Keplerian frequency is $\sim 10^{-2}$ Hz]. 
The cosines of endpoint tilt angles $\cos{\theta_{1,2}^{\rm ude}}$ 
are only exact in the unphysical limit of zero separation and are obtained using $2$PN spin evolution equations. However, even when starting at a reference frequency of $20$~Hz, we find that cosines of the injected tilts at infinity obtained with the hybrid evolution (using $3$PN spin evolution equations) are very close to the exact up-down configuration ($\cos\theta_1 = 1$, $\cos\theta_2 = -1$). Specifically, the cosines of the injected tilts at infinity differ from the exact up-down configuration by at most $1\%$ ($2\%$) for the $100, 200 M_{\odot}$ ($20, 50 M_{\odot}$) cases.

\begin{figure*}[h]
\centering
\subfloat{
\includegraphics[width=0.95\linewidth]{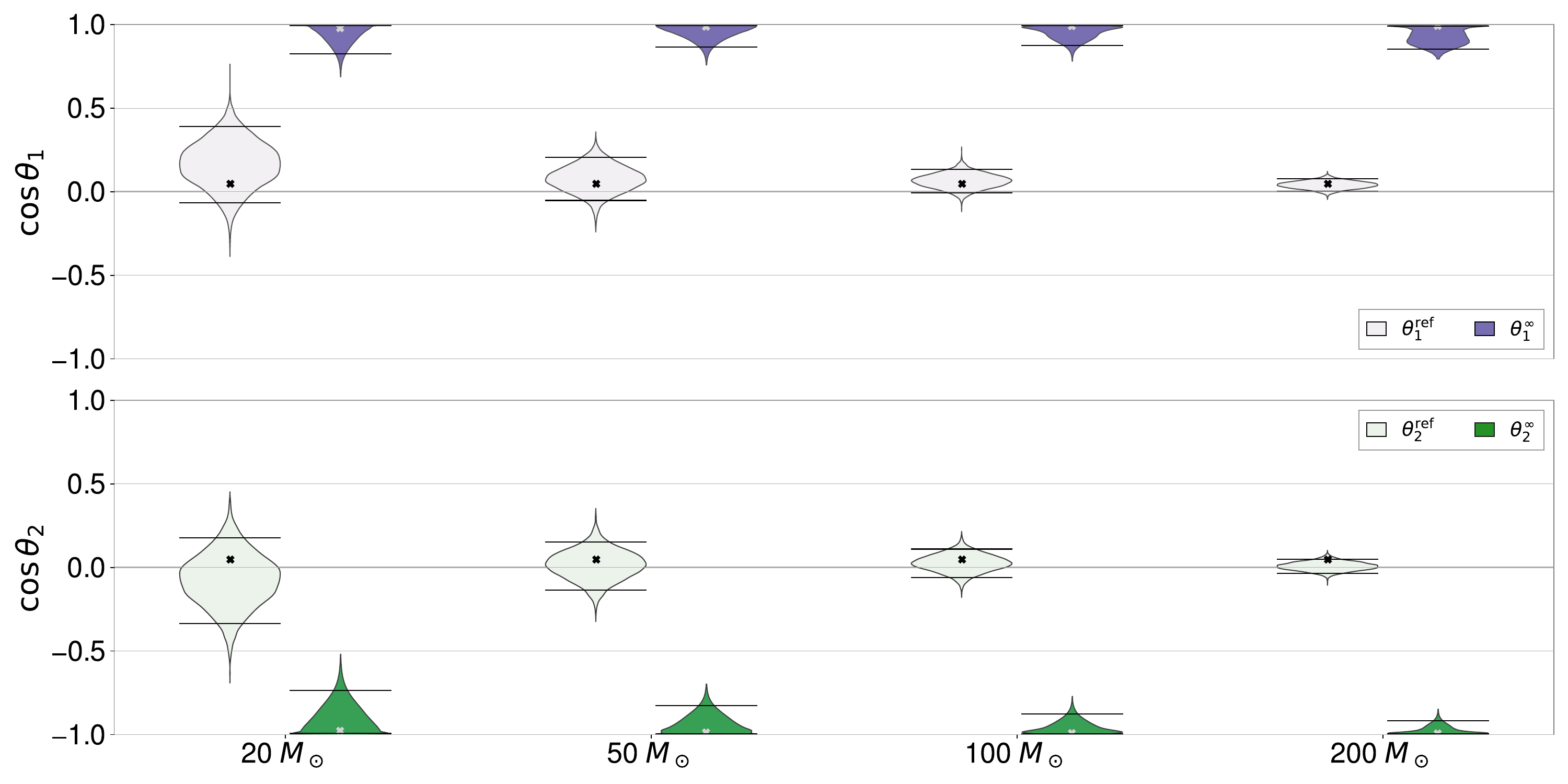}
}
\caption{\label{fig:updown_endpoints} Comparison of posterior distributions of tilts injected at the endpoints of the up-down instability at the reference frequency and infinity. We find that all the posteriors of the tilts at infinity rail against the prior boundary of $\cos{\theta} = \pm 1$.
}
\end{figure*}

Figure~\ref{fig:updown_endpoints} shows the inferred tilt posteriors at $f_{\rm ref}$ and infinity for all four cases. The tilt endpoints, which are close to being in-plane at the reference frequency, move to the up-down configuration at infinite separation. This shows that we would be able to identify such a binary as an up-down instability configuration with high confidence in the A+/AdV+ era. (See~\cite{DeRenzis:2023lwa} for a complementary study using the O4 sensitivity that also performs Bayesian model selection.) The result also complements our previous results, where the tilts were often in an up-down configuration at $f_\text{ref}$ and were closer to in-plane at infinite separation.
The SNRs for the $20$, $50$, $100$, and $200 M_{\odot}$ cases are $46$, $98$, $131$, and $228$ respectively. While the inferred tilts are more well-constrained for the high-mass, high-SNR binaries, the tilts at infinity clearly favor the up-down configuration in all cases.

\subsection{Hybrid vs. Precession-averaged Evolution}
\label{subsec:hyb_vs_prec}

\begin{figure*}[!htb]
\centering
\subfloat{
\includegraphics[width=0.95\linewidth]{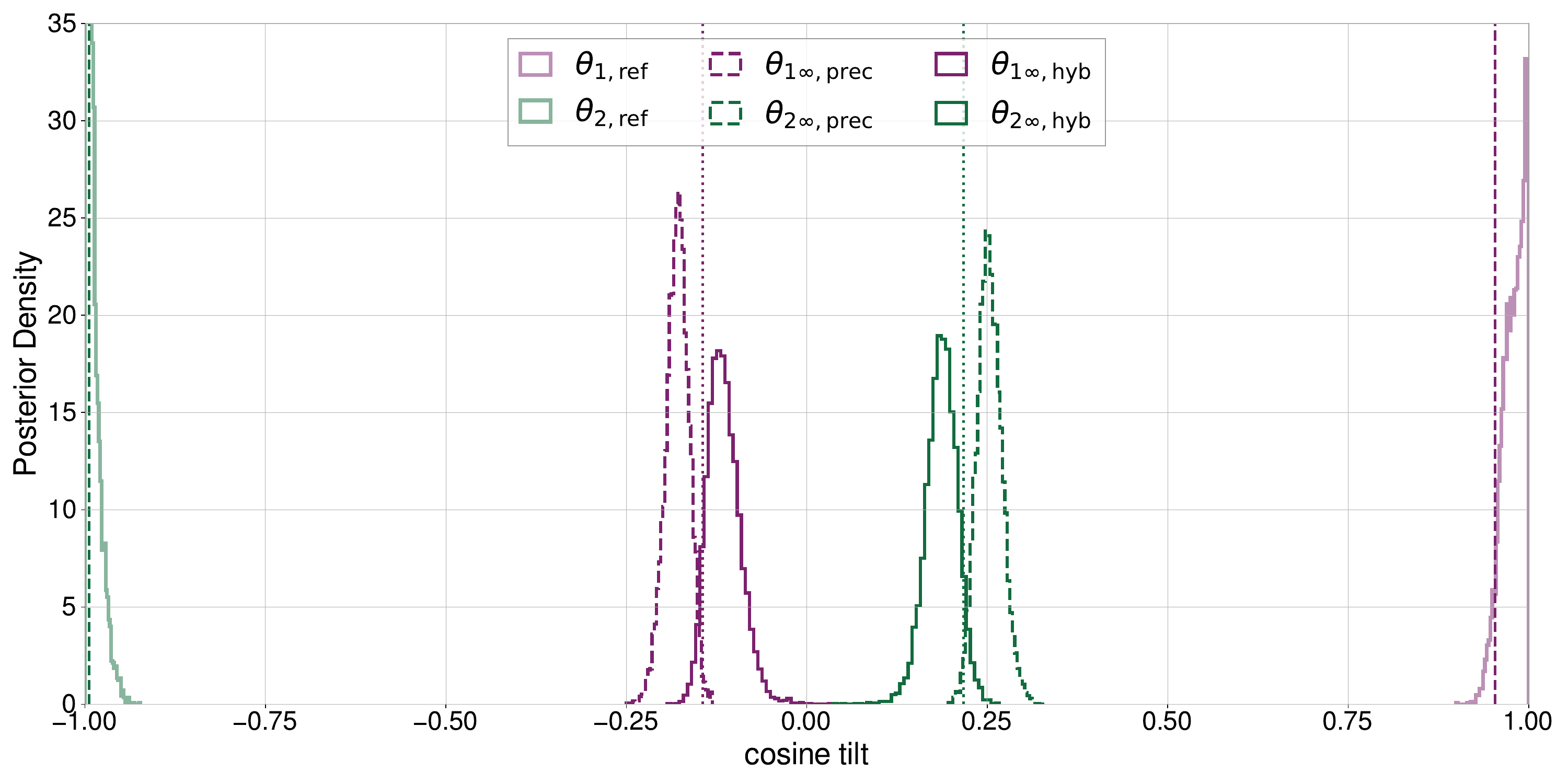}
}
\caption{\label{fig:hyb_vs_prec} Comparison of tilt posteriors at infinity computed using hybrid and precession-only evolution methods for the $M=200 M_\odot$, $q=1/1.1$, $\chi_1=\chi_2=0.95$, $\phi_{12} = \pi$ case. We also show the tilt posteriors at $f_{\rm ref}$ to emphasize the large difference compared to both calculations of the tilts at infinity. The dashed (dotted) vertical lines represent injected tilt values at $f_{\rm ref}$ (infinity, using hybrid evolution). 
}
\end{figure*}

For certain binaries in our set of simulations as well as the morphology injections, we observe  significant differences in tilt distributions at infinity computed using two different spin evolution approaches: the purely precession-averaged spin evolution and the hybrid orbit-averaged + precession-averaged evolution. Figure~\ref{fig:hyb_vs_prec} illustrates this for one case that shows the largest deviation in the median of the distributions, for the binary with $\{M=200 M_{\odot}, q=1/1.1, \chi_1=\chi_2=0.95, \phi_{12} = \pi\}$. As discussed in~\cite{Johnson-McDaniel:2021rvv}, the precession-averaged evolution only includes the leading PN spin-orbit and spin-spin effects (i.e., up to $2$PN) in the evolution, while the hybrid method includes higher orders (up to $3$PN in spin effects) in the orbit-averaged evolution that is used close to merger. This makes the hybrid evolution code more accurate, particularly for high mass systems where the binary is already in the strong-field regime at the reference frequency making the lower-PN expressions (and the precession-averaged approximation itself) insufficiently accurate to describe the binary's dynamics. Of course, we find even larger differences between the hybrid evolution and precession-averaged evolution for injected values in the cases with very high sensitivity to the parameters discussed in Sec.~\ref{sec:inj-study} and Appendix~\ref{sec:appdxA}. However, since this large difference only occurs for very precise binary parameters, it does not translate to a significant difference in the posteriors obtained with the two different evolution methods.

For the injections with different spin morphologies, we observe similar shifts in the hybrid and precession-only distributions as above for the L$0$ boundary case with high spins. Interestingly, these deviations are seen for both the total mass ($20M_\odot$, $75M_\odot$) injections, showing that it is necessary to use hybrid evolution to get accurate results even for lower-mass binaries, where one might not expect the effects of the hybrid evolution to be as significant. Besides a very slight deviation in the hybrid and precession-averaged distributions at infinity for the C-L$0$ morphology, the results of the two evolutions are indistinguishable for all other morphologies. 
For injections involving the endpoints of the up-down instability, we see no differences between hybrid and precession-averaged evolution.

\section{Conclusions}
\label{sec:conclusion}

Binary black hole spin tilts can serve as proxies to determine and distinguish between different formation channels of such systems~\cite{Vitale:2015tea,Rodriguez:2016vmx,Farr:2017uvj,Farr:2017gtv}. However, the tilts close to the binary's merger (where they are traditionally estimated in gravitational wave data analysis) can differ significantly from those at formation, which are well approximated by the tilts at infinite separation in many cases---see~\cite{Johnson-McDaniel:2021rvv}. The latest LVK analyses~\cite{LIGOScientific:2021usb,LIGOScientific:2021djp} quote the tilts at infinity, though the tilts are sufficiently uncertain that there is not a significant different between the tilts at the reference frequency close to merger and at infinity, as discussed in~\cite{Johnson-McDaniel:2021rvv}.
However, plus-era detectors~\cite{KAGRA:2013rdx} promise to constrain the tilts well enough (at least for events with high enough SNR) that the inferred posteriors on these parameters they will be significantly different when evolved back in time to infinite separation in many cases.
In this paper, we determined what kind of BBH configurations will show the largest difference in spin tilts between the reference frequency and at infinite separation. Further, we performed an injection study to explore the evolution of the inferred spin tilt posteriors to infinite separation for such binaries.

We found that BBHs with more equal masses regardless of their spin magnitudes and for a given mass ratio BBHs that have higher spins show large difference in tilts at reference frequency and infinity. We then simulated GW signals in the LIGO and Virgo detectors with A+/AdV+ sensitivity (and zero noise). These injections have three total masses \{$50M_\odot$, $100M_\odot$, $200M_\odot$\}, 
three mass ratios ($1/1.1$, $1/3$, $1/8$), and two spin magnitudes ($0.5$, $0.95$). The spin angles $\theta_1$, $\theta_2$, $\phi_{12}$ at $f_{\rm ref}$ for each of these injections were chosen to give the maximum difference in tilts at the reference frequency and infinite separation. We used the \textsc{IMRPhenomXPHM} waveform model~\cite{Pratten:2020ceb} to create the injections as well as for the recovery template in our parameter estimation. We evolve the posterior samples from the reference frequency to infinity using our hybrid evolution code~\cite{Johnson-McDaniel:2021rvv} to obtain the inferred tilts at infinity. We found that for the majority of our injections the 90\% CIs of tilt posteriors at the reference frequency and infinite separation overlap, though there are still noticeable shifts in most cases. However, in some cases the 90\% CIs are completely disjoint and there are even cases where the two distributions are very well separated. This is particularly the case when the BBHs are highly symmetric in mass ($q=1/1.1$) and spin magnitudes ($\chi_1=\chi_2$) as well as spinning very fast ($\chi_1, \chi_2=0.95$). 

We also considered the results in~\cite{Johnson-McDaniel:2023oea}, which performed parameter estimation on simulated injections that lie in various precessional morphologies, and explored the behavior at infinite separation. We observed significant differences in tilt posteriors for some of these cases even though they had not been selected specifically for large tilt differences. Finally, we injected BBHs at the endpoint of the up-down instability at 20 Hz and found that we indeed recover tilt posteriors at infinite separation that clearly prefer the up-down configuration of $\cos{\theta_{1}} = 1$, $\cos{\theta_{2}} = -1$.    

These differences between the tilts at the reference frequency and infinity will be even more important for third-generation ground-based detectors such as Cosmic Explorer~\cite{Evans:2021gyd} and Einstein Telescope~\cite{Hild:2010id}, where both tilts at the reference frequency may be measured with $90\%$ credible intervals of $0.02$ (we only obtain such accuracy for the primary tilt in the $q = 1/8$ case, while the best accuracy obtained in the $q = 1/1.1$ cases is $0.06$). The same is true for space-based detectors such as LISA~\cite{LISA}, where the tilts at the reference frequency may be measured with $90\%$ credible intervals with widths of $0.003$ (for the primary tilt) in some cases~\cite{Pratten:2023krc}. In these cases, it will likely be important to improve upon the accuracy of the PN evolution, as we have started to do by using the surrogate spin evolution in its region of availability before switching to the hybrid evolution. The change in the posteriors due to using surrogate evolution is barely discernible even for the case of high spin binaries and the plus-era posteriors we consider, but could prove to be significant in other cases or in third-generation detectors. In particular, Ref.~\cite{Johnson-McDaniel:2021rvv} finds that individual samples can have differences in the cosines of the tilts as large as $\sim 1$ due to the surrogate evolution.

There are also possibilities for improving the speed of the calculation, either by creating a surrogate model or by averaging the tilts at infinity over a cycle of oscillations, as discussed in~\cite{Johnson-McDaniel:2021rvv}, and/or using the more efficient orbit-averaged evolution given in~\cite{Yu:2023lml}. There is also an improved precession-averaged formalism given in~\cite{Gerosa:2023xsx} that might improve efficiency and/or accuracy in some cases, particularly for close-to-equal-mass binaries. Finally, while the current calculation is only applicable to binary black holes, since the precession-averaged evolution is restricted to this case (it is only applicable to the case where both objects have black hole spin-induced quadrupoles), there is recent work~\cite{LaHaye:2022yxa} that should allow one to extend the precession-averaged evolution to binaries involving neutron stars.

\section*{Acknowledgments}
We thank Serguei Ossokine for carefully reading the manuscript and providing helpful comments and the anonymous referee for useful suggestions. NKJ-M is supported by the NSF grant AST-2205920. AG is supported in part by NSF grants AST-2205920 and PHY-2308887. KSP acknowledges support from the Dutch Research Council (NWO). NVK is thankful to the Max Planck Society's Independent Research Group Grant, Science and Engineering Research Board National Post Doctoral Fellowship (N-PDF).
The authors are grateful for computational resources provided by the LIGO Lab and supported by NSF Grants PHY-0757058 and PHY-0823459. We also acknowledge the use of the Maple cluster at the University of Mississippi (funded by NSF Grant CHE-1338056), the IUCAA LDG cluster Sarathi, the University of Birmingham's BlueBEAR HPC service, Nikhef's Visar cluster, and Max Planck Computing and Data Facility's clusters Raven and Cobra for the computational/numerical work.

This research has made use of Parallel Bilby v1.1.3~\cite{pbilby_paper, bilby_paper}, a parallelized Bayesian inference Python package, and Dynesty v1.1~\cite{dynesty_paper, skilling2004, skilling2006}, a nested sampler, to perform Bayesian parameter estimation. The software packages AstroPy~\cite{Astropy2018}, LALSuite~\cite{lalsuite}, Matplotlib~\cite{plt4160265}, NumPy~\cite{Harris:2020xlr}, Pandas~\cite{pandas_paper}, PESummary~\cite{Hoy:2020vys}, SciPy~\cite{2020SciPy-NMeth}, and Seaborn~\cite{Waskom2021} were utilised for data analysis. This is LIGO document P2300128.

\appendix

\section{Sensitivity to inputs}
\label{sec:appdxA}
 
\begin{table}[t] 
\centering
\setlength{\tabcolsep}{3pt} 
\renewcommand{\arraystretch}{2} 
\begin{tabular}{ccccc} 
\hline 
$M (M_{\odot})$ & $\chi_{1}$ & $\chi_{2}$ & $\theta_{1}^{\rm ref}$ & $\theta_{2}^{\rm ref}$ \\ 
\hline \hline 
50 & 0.5 & 0.5 & 1.8061875622062771 & 1.2040619984276995 \\ 
\hline 
50 & 0.5 & 0.95 & 2.1901356745982299 & 1.6547753106912710 \\ 
\hline 
50 & 0.95 & 0.95 & 1.8453795275676992 & 1.1595409673747283 \\ 
\hline 
50 & 0.95 & 0.5 & 1.4023374949509382 & 0.7349404682555843 \\ 
\hline 
100 & 0.5 & 0.5 & 1.8295586677990303 & 1.1769010782793288 \\ 
\hline 
100 & 0.5 & 0.95 & 2.2116349548106866 & 1.6445107009602780 \\ 
\hline 
100 & 0.95 & 0.95 & 1.8475321288385302 & 1.1568255975485275 \\ 
\hline 
100 & 0.95 & 0.5 & 1.3528117893583296 & 0.8760247525344458 \\ 
\hline 
200 & 0.5 & 0.5 & 1.8081591435948825 & 1.2010486593621947 \\ 
\hline 
200 & 0.5 & 0.95 & 2.2505268577289996 & 1.6262244552266782 \\ 
\hline 
200 & 0.95 & 0.95 & 1.8503377465477728 & 1.1529704131721257 \\ 
\hline 
200 & 0.95 & 0.5 & 1.3794912619875150 & 0.8012609687138224 \\ 
\hline 
\end{tabular} 

\caption{\label{table:hypersensitive_params} %
The tilts at $f_{\rm ref}$ to $16$ decimal places 
for the $12$ comparable mass ($q=0.9090909090909091 \simeq 1/1.1$) cases with $\phi_{12}=0$ for which the output of tilts at infinity using hybrid evolution is hypersensitive to the input parameters. Specifically, this gives the cases for which there is a difference of greater than $0.01$ in either of the tilts at infinity when comparing the values obtained using the values given here for the tilts at $f_{\rm ref}$ and those obtained after first rounding the tilts at $f_{\rm ref}$ to two decimal places, as in Tables~\ref{table:params_eqspins} and~\ref{table:params_uneqspins}. These differences in the cases given here range from $0.06$ to $0.70$.
}
\end{table}

In this section, we provide details that describe the sensitivity to precise input values when evolving the tilt configurations optimized for large differences between the tilts at the reference frequency and at infinity in some cases where $\phi_{12}=0$. We give the precise input parameters for these cases in Table~\ref{table:hypersensitive_params}.
We use the $M=200 M_{\odot}$, $q=1/1.1$, $\chi_1=\chi_2=0.5$ binary to illustrate this effect. Using the default settings of the hybrid evolution code ($3$PN-order spin terms and SpinTaylorT5 approximant~\cite{SpinTaylor_TechNote}), and setting $\phi_{12}=0$, the optimization code returns tilts of $\theta_1^{\rm ref}=1.8081591435948825$, $\theta_2^{\rm ref}=1.2010486593621947$ at $f_{\rm ref}=10$ Hz. For these precise inputs, the expected tilts at infinity given by the hybrid code, again with default settings, are (rounded to three decimal digits): \{$\theta_1^{\infty}=0.007$, $\theta_2^{\infty}=3.134$\},
very close to the exact up-down configuration.\footnote{Exactly how close this output is to the up-down configuration may change somewhat (by as much as $0.07$ in our checks) with the software version(s) used, presumably due to  different orders of floating-point operations therein.} Since the up-down configuration is unstable, this instability presumably explains the sensitivity to the input parameters. For instance, when the input tilts are rounded off to $6$ decimal places: \{$\theta_1^{\rm ref}=1.808159$, $\theta_2^{\rm ref}=1.201049$\}, the hybrid evolution output gives \{$\theta_1^{\infty}=0.471$, $\theta_2^{\infty}=2.647$\}. With further rounding off to just one decimal place: \{$\theta_1^{\rm ref}=1.8$, $\theta_2^{\rm ref}=1.2$\}, the output becomes \{$\theta_1^{\infty}=0.671$, $\theta_2^{\infty}=2.437$\}. 
We have checked that the results with the full accuracy of the inputs are not changed significantly when using more stringent settings such as smaller sampling intervals used in interpolating the final output of the orbit-averaged evolution, or making small changes in the transition frequency which switches from orbit to precession-averaged evolution, though as mentioned above, they are apparently sensitive to different floating-point operation ordering.

Additionally, we also checked the effect of tightening the tolerance settings of the integrator used in the orbit-averaged evolution by two orders of magnitude. The tilts at infinity for the same case as above with these settings are \{$\theta_1^{\infty}=0.349$, $\theta_2^{\infty}=2.775$\} (rounded to three decimal places). Even though these results are notably different from the close-to-up-down configuration found with the looser tolerance, they still have a significant sensitivity to the precise input values. In particular, the results when truncating to $6$ decimal places are the same as with the looser default tolerance to the number of decimal places quoted above. We also find that with the tighter tolerance, one obtains tilts at infinity that are even closer to the up-down configuration with a small perturbation of the initial values (obtained by performing the optimization with the tighter integrator tolerance), i.e., \{$\theta_1^{\rm ref}=1.8081591424304095$, $\theta_2^{\rm ref}=1.2010486608642275 $\} gives \{$\theta_1^{\infty}=0.003$, $\theta_2^{\infty}=3.138$\}. However, the magnitudes of the differences in the cosines of these tilts at infinity and the ones obtained with the looser tolerance are $\sim 2\times 10^{-5}$, well below the $\sim 10^{-4}$ accuracy of the hybrid evolution due to the choice of the transition frequency shown in Fig.~6 of~\cite{Johnson-McDaniel:2021rvv}. 

Thus, the parameter estimation results in this paper will only have negligible differences due to using the tighter tolerance, since the injected parameters would only differ past the seventh decimal place, and the individual samples are not close enough to the unstable point for the different tolerance to cause significant differences: The largest difference in the cosines of the tilts at infinity samples is $< 10^{-6}$.
We also checked that tightening the integrator tolerance or using a higher-order integration method in the precession-averaged part of the hybrid evolution did not lead to significant differences in the results---the maximum difference between the cosines of the tilts at infinity (considering both tilts) is $2.4\times10^{-7}$. When studying these particular unstable cases, it would likely be worthwhile to use arbitrary precision for the orbit-averaged evolution. However, as discussed above, such refinements to the hybrid evolution are not expected to be necessary for parameter estimation applications, even for third-generation or space-based GW detectors (see the expected measurement accuracies given in Sec.~\ref{sec:conclusion}). 

This sensitivity of the tilts at infinity to the input values in these cases is present for all input parameters, not just restricted the input tilts. For instance, when the mass ratio is changed from $q=0.9090909090909091$ (the decimal approximation to $q = 1/1.1$ we used) to $q=0.9$, the tilts at infinity change to \{$\theta_1^{\infty}=0.740$, $\theta_2^{\infty}=2.379$\}. This sensitivity also extends to the precise values of conversion quantities used in generating the input parameters, such as the solar mass in SI units (used to convert component masses). 
The value of the solar mass used in our evolutions is $1M_\odot = 1.9884099021470415 \times 10^{30}$~kg, which should be used to reproduce our results.\footnote{The value for this quantity in LALSuite~\cite{lalsuite} (MSUN\_SI) has recently been updated to a slightly different value.}
Using only the precession-averaged evolution with the untruncated tilt inputs also gives tilts at infinity values of \{$\theta_1^{\infty}=0.702$, $\theta_2^{\infty}=2.399$\} which are closer to the results obtained with truncated inputs. When using spin terms only to $2.5$PN ($2$PN), the tilts at infinity obtained are \{$\theta_1^{\infty}=0.673~(0.687)$, $\theta_2^{\infty}=2.434~(2.416)$\}, showing how these approach the values from only precession-averaged evolution more closely as one reduces the spin order. Similarly, keeping the $3$PN spin terms but changing the approximant to SpinTaylorT1 (SpinTaylorT4) 
gives tilts at infinity of \{$\theta_1^{\infty}=0.641~(0.574)$, $\theta_2^{\infty}=2.467~(2.539)$\}.

Further, as mentioned in Sec.~\ref{sec:inj-study}, we get slightly different peak tilt configurations when running the optimization code using a different approximant in the hybrid evolution, but  the hypersensitive behavior outlined above is also seen for these configurations. For example, when optimizing using SpinTaylorT4 and $3$PN spin terms for the same binary, the peak tilt configuration is \{$\theta_{1}^{\rm ref} = 1.8712719865486664$, $\theta_{2}^{\rm ref} = 1.1278607980431046$\}, giving tilts at infinity of \{$\theta_{1}^{\infty} = 0.153$, $\theta_{2}^{\infty} = 2.980$\}. Using $2.5$PN spin terms with the SpinTaylorT5 approximant also gives very similar results. 
When optimized using the default SpinTaylorT5 but with $2$PN spin terms on the other hand, the optimum tilt configuration becomes \{$\theta_{1}^{\rm ref} = 1.7862445760631980$, $\theta_{2}^{\rm ref} = 1.2290171723775418$\} with tilts at infinity of \{$\theta_{1}^{\infty} = 0.009$, $\theta_{2}^{\infty} = 3.131$\}.

\section{Tabulated Results}
\label{sec:appdxB}

Here we give the values for the injected intrinsic parameters for our simulations with the spin angles chosen to maximize the difference between the tilts at infinity and the reference frequency (as outlined in Sec.~\ref{sec:inj-study}). We also show the median and 90\% CI bounds of their recovered posteriors along with the median and 90\% CI bounds for the spin tilt posteriors at infinity. Specifically, Table~\ref{table:params_eqspins} provides these for equal-spin binaries and Table~\ref{table:params_uneqspins} for unequal spin binaries.

\begin{table*}[] 
\centering
\renewcommand{\arraystretch}{2} 
\begin{tabular}{ccccccccccccc} 
\hline 
\multicolumn{5}{c|}{Injected} & \multicolumn{8}{|c}{Recovered} \\ 
\hline 
$q$ & $\chi_{1}$ & $\chi_{2}$ & $\theta_{1}^{\rm ref}$ & $\theta_{2}^{\rm ref}$ & $M (M_{\odot})$ & $q$ & $\chi_{1}$ & $\chi_{2}$ & $\theta_{1}^{\rm ref}$ & $\theta_{2}^{\rm ref}$ & $\theta_{1}^{\infty}$ & $\theta_{2}^{\infty}$  \\ 
\hline \hline 
\multicolumn{13}{c}{$M = 50 M_{\odot}$} \\ 
\hline \hline 
1/8 & 0.5 & 0.5 & 1.41 & 1.65 \vline \vline & $50.1_{-1.1}^{+1.2}$ & $0.125_{-0.006}^{+0.006}$ & $0.50_{-0.01}^{+0.02}$ & $0.27_{-0.24}^{+0.47}$ & $1.42_{-0.05}^{+0.06}$ & $1.55_{-0.97}^{+0.91}$ & $1.42_{-0.05}^{+0.05}$ & $1.49_{-0.93}^{+0.95}$ \\ 
\hline 
1/8 & 0.95 & 0.95 & 1.27 & 1.73 \vline \vline & $50.0_{-1.1}^{+1.0}$ & $0.125_{-0.005}^{+0.006}$ & $0.94_{-0.02}^{+0.02}$ & $0.72_{-0.50}^{+0.24}$ & $1.27_{-0.03}^{+0.03}$ & $1.76_{-0.44}^{+0.42}$ & $1.28_{-0.03}^{+0.03}$ & $1.59_{-0.43}^{+0.47}$ \\ 
\hline 
1/3 & 0.5 & 0.5 & 1.36 & 1.71 \vline \vline & $50.20_{-0.85}^{+0.95}$ & $0.33_{-0.02}^{+0.02}$ & $0.49_{-0.03}^{+0.03}$ & $0.55_{-0.18}^{+0.20}$ & $1.31_{-0.10}^{+0.11}$ & $1.86_{-0.36}^{+0.35}$ & $1.35_{-0.11}^{+0.11}$ & $1.73_{-0.34}^{+0.36}$ \\ 
\hline 
$^{\dag}$1/3 & 0.95 & 0.95 & 1.28 & 1.52 \vline \vline & $49.89_{-0.64}^{+0.64}$ & $0.34_{-0.01}^{+0.01}$ & $0.94_{-0.02}^{+0.02}$ & $0.93_{-0.10}^{+0.05}$ & $1.26_{-0.04}^{+0.04}$ & $1.60_{-0.13}^{+0.14}$ & $1.16_{-0.04}^{+0.04}$ & $1.90_{-0.12}^{+0.13}$ \\ 
\hline 
$^{\dag}$1/1.1 & 0.5 & 0.5 & 1.81 & 1.20 \vline \vline & $49.91_{-0.13}^{+0.14}$ & $0.95_{-0.06}^{+0.05}$ & $0.24_{-0.20}^{+0.51}$ & $0.64_{-0.52}^{+0.28}$ & $1.60_{-0.67}^{+0.92}$ & $1.50_{-0.42}^{+0.40}$ & $1.1_{-0.7}^{+1.3}$ & $1.70_{-0.32}^{+0.88}$ \\ 
\hline 
1/1.1 & 0.5 & 0.5 & 0.54 & 2.60 \vline \vline & $49.94_{-0.15}^{+0.21}$ & $0.89_{-0.05}^{+0.03}$ & $0.54_{-0.09}^{+0.09}$ & $0.57_{-0.14}^{+0.15}$ & $0.57_{-0.35}^{+0.56}$ & $2.65_{-0.58}^{+0.30}$ & $1.40_{-0.12}^{+0.15}$ & $1.74_{-0.23}^{+0.14}$ \\ 
\hline 
$^{\dag}$1/1.1 & 0.95 & 0.95 & 1.85 & 1.16 \vline \vline & $49.99_{-0.11}^{+0.13}$ & $0.92_{-0.03}^{+0.06}$ & $0.91_{-0.10}^{+0.07}$ & $0.93_{-0.11}^{+0.05}$ & $1.80_{-0.14}^{+0.09}$ & $1.21_{-0.09}^{+0.14}$ & $0.65_{-0.17}^{+0.15}$ & $2.40_{-0.17}^{+0.20}$ \\ 
\hline 
1/1.1 & 0.95 & 0.95 & 0.35 & 2.81 \vline \vline & $50.02_{-0.11}^{+0.12}$ & $0.912_{-0.006}^{+0.011}$ & $0.94_{-0.06}^{+0.04}$ & $0.94_{-0.09}^{+0.05}$ & $0.31_{-0.09}^{+0.09}$ & $2.82_{-0.10}^{+0.09}$ & $1.64_{-0.03}^{+0.03}$ & $1.40_{-0.04}^{+0.03}$ \\ 
\hline 
\multicolumn{13}{c}{$M = 100 M_{\odot}$} \\ 
\hline \hline 
1/8 & 0.5 & 0.5 & 1.36 & 1.67 \vline \vline & $100.2_{-1.1}^{+1.1}$ & $0.124_{-0.003}^{+0.003}$ & $0.94_{-0.01}^{+0.01}$ & $0.86_{-0.23}^{+0.12}$ & $1.36_{-0.02}^{+0.02}$ & $1.63_{-0.27}^{+0.21}$ & $1.39_{-0.02}^{+0.02}$ & $1.38_{-0.27}^{+0.21}$ \\ 
\hline 
1/8 & 0.95 & 0.95 & 1.18 & 1.76 \vline \vline & $99.94_{-0.95}^{+0.97}$ & $0.125_{-0.002}^{+0.002}$ & $0.942_{-0.009}^{+0.007}$ & $0.89_{-0.19}^{+0.09}$ & $1.18_{-0.02}^{+0.02}$ & $1.82_{-0.20}^{+0.19}$ & $1.21_{-0.02}^{+0.02}$ & $1.57_{-0.20}^{+0.20}$ \\ 
\hline 
1/3 & 0.5 & 0.5 & 1.30 & 1.74 \vline \vline & $100.17_{-0.86}^{+0.84}$ & $0.330_{-0.008}^{+0.009}$ & $0.48_{-0.02}^{+0.02}$ & $0.40_{-0.16}^{+0.16}$ & $1.29_{-0.07}^{+0.08}$ & $1.79_{-0.38}^{+0.38}$ & $1.33_{-0.08}^{+0.08}$ & $1.63_{-0.37}^{+0.38}$ \\ 
\hline 
$^{\dag}$1/3 & 0.95 & 0.95 & 1.24 & 1.52 \vline \vline & $100.01_{-0.64}^{+0.61}$ & $0.332_{-0.006}^{+0.007}$ & $0.95_{-0.01}^{+0.01}$ & $0.94_{-0.07}^{+0.05}$ & $1.24_{-0.03}^{+0.03}$ & $1.55_{-0.11}^{+0.10}$ & $1.12_{-0.03}^{+0.04}$ & $1.89_{-0.12}^{+0.10}$ \\ 
\hline 
$^{\dag}$1/1.1 & 0.5 & 0.5 & 1.83 & 1.18 \vline \vline & $99.96_{-0.46}^{+0.41}$ & $0.91_{-0.03}^{+0.03}$ & $0.36_{-0.24}^{+0.25}$ & $0.62_{-0.29}^{+0.30}$ & $1.86_{-0.17}^{+0.75}$ & $1.30_{-0.24}^{+0.13}$ & $1.04_{-0.24}^{+0.93}$ & $1.86_{-0.40}^{+0.39}$ \\ 
\hline 
1/1.1 & 0.5 & 0.5 & 0.55 & 2.59 \vline \vline & $100.09_{-0.71}^{+0.53}$ & $0.91_{-0.02}^{+0.01}$ & $0.50_{-0.07}^{+0.05}$ & $0.51_{-0.06}^{+0.06}$ & $0.55_{-0.13}^{+0.15}$ & $2.55_{-0.18}^{+0.13}$ & $1.53_{-0.10}^{+0.09}$ & $1.52_{-0.09}^{+0.14}$ \\ 
\hline 
$^{\dag}$1/1.1 & 0.95 & 0.95 & 1.85 & 1.16 \vline \vline & $99.87_{-0.42}^{+0.43}$ & $0.91_{-0.02}^{+0.02}$ & $0.95_{-0.06}^{+0.04}$ & $0.95_{-0.07}^{+0.04}$ & $1.84_{-0.06}^{+0.05}$ & $1.18_{-0.06}^{+0.07}$ & $0.70_{-0.11}^{+0.12}$ & $2.42_{-0.09}^{+0.10}$ \\ 
\hline 
1/1.1 & 0.95 & 0.95 & 0.36 & 2.89 \vline \vline & $100.01_{-0.43}^{+0.43}$ & $0.911_{-0.004}^{+0.006}$ & $0.93_{-0.04}^{+0.04}$ & $0.95_{-0.05}^{+0.03}$ & $0.28_{-0.10}^{+0.10}$ & $2.91_{-0.09}^{+0.08}$ & $1.70_{-0.02}^{+0.02}$ & $1.37_{-0.03}^{+0.03}$ \\ 
\hline 
\multicolumn{13}{c}{$M = 200 M_{\odot}$} \\ 
\hline \hline 
1/8 & 0.5 & 0.5 & 1.30 & 1.68 \vline \vline & $199.8_{-2.4}^{+2.5}$ & $0.125_{-0.003}^{+0.003}$ & $0.50_{-0.02}^{+0.02}$ & $0.44_{-0.26}^{+0.24}$ & $1.31_{-0.04}^{+0.04}$ & $1.67_{-0.68}^{+0.69}$ & $1.32_{-0.04}^{+0.05}$ & $1.52_{-0.67}^{+0.70}$ \\ 
\hline 
1/8 & 0.95 & 0.95 & 1.18 & 1.76 \vline \vline & $199.73_{-0.80}^{+0.75}$ & $0.125_{-0.001}^{+0.001}$ & $0.946_{-0.005}^{+0.004}$ & $0.91_{-0.11}^{+0.07}$ & $1.18_{-0.01}^{+0.01}$ & $1.85_{-0.15}^{+0.16}$ & $1.21_{-0.01}^{+0.01}$ & $1.54_{-0.15}^{+0.16}$ \\ 
\hline 
$^{\dag}$1/3 & 0.5 & 0.5 & 1.38 & 1.53 \vline \vline & $199.6_{-1.4}^{+1.3}$ & $0.334_{-0.008}^{+0.008}$ & $0.50_{-0.03}^{+0.03}$ & $0.48_{-0.13}^{+0.16}$ & $1.42_{-0.09}^{+0.10}$ & $1.42_{-0.50}^{+0.37}$ & $1.36_{-0.10}^{+0.11}$ & $1.63_{-0.51}^{+0.37}$ \\ 
\hline 
$^{\dag}$1/3 & 0.95 & 0.95 & 1.16 & 1.50 \vline \vline & $200.1_{-1.1}^{+1.2}$ & $0.334_{-0.004}^{+0.005}$ & $0.951_{-0.008}^{+0.010}$ & $0.93_{-0.07}^{+0.05}$ & $1.16_{-0.02}^{+0.02}$ & $1.51_{-0.09}^{+0.07}$ & $1.01_{-0.03}^{+0.03}$ & $1.94_{-0.09}^{+0.07}$ \\ 
\hline 
$^{\dag}$1/1.1 & 0.5 & 0.5 & 1.81 & 1.20 \vline \vline & $200.2_{-1.1}^{+1.1}$ & $0.91_{-0.02}^{+0.02}$ & $0.41_{-0.26}^{+0.39}$ & $0.64_{-0.42}^{+0.32}$ & $1.86_{-0.18}^{+0.47}$ & $1.29_{-0.56}^{+0.13}$ & $1.11_{-0.34}^{+0.38}$ & $1.83_{-0.30}^{+0.47}$ \\ 
\hline 
1/1.1 & 0.5 & 0.5 & 0.55 & 2.60 \vline \vline & $199.9_{-1.1}^{+1.1}$ & $0.91_{-0.01}^{+0.01}$ & $0.47_{-0.07}^{+0.06}$ & $0.50_{-0.06}^{+0.06}$ & $0.41_{-0.25}^{+0.28}$ & $2.65_{-0.20}^{+0.24}$ & $1.45_{-0.14}^{+0.09}$ & $1.64_{-0.10}^{+0.17}$ \\ 
\hline 
$^{\dag}$1/1.1 & 0.95 & 0.95 & 1.85 & 1.15 \vline \vline & $199.93_{-0.93}^{+0.98}$ & $0.91_{-0.01}^{+0.02}$ & $0.95_{-0.04}^{+0.03}$ & $0.95_{-0.05}^{+0.04}$ & $1.84_{-0.03}^{+0.03}$ & $1.18_{-0.04}^{+0.04}$ & $0.67_{-0.08}^{+0.10}$ & $2.46_{-0.08}^{+0.07}$ \\ 
\hline 
1/1.1 & 0.95 & 0.95 & 0.31 & 3.04 \vline \vline & $200.6_{-1.1}^{+1.0}$ & $0.910_{-0.005}^{+0.007}$ & $0.94_{-0.05}^{+0.04}$ & $0.96_{-0.05}^{+0.03}$ & $0.20_{-0.14}^{+0.12}$ & $3.02_{-0.13}^{+0.08}$ & $1.69_{-0.04}^{+0.03}$ & $1.38_{-0.03}^{+0.04}$ \\ 
\hline 
\end{tabular} 

\caption{\label{table:params_eqspins} Injected and recovered parameters for equal-spin injections. Rows marked by a $(${\dag}$)$ have the injected value of $\phi_{12}$ set to $0$, while it is $\pi$ for the unmarked cases. 
}
\end{table*}

\begin{table*}[] 
\centering
\renewcommand{\arraystretch}{2} 
\begin{tabular}{ccccccccccccc} 
\hline 
\multicolumn{5}{c|}{Injected} & \multicolumn{8}{|c}{Recovered} \\ 
\hline 
$q$ & $\chi_{1}$ & $\chi_{2}$ & $\theta_{1}^{\rm ref}$ & $\theta_{2}^{\rm ref}$ & $M (M_{\odot})$ & $q$ & $\chi_{1}$ & $\chi_{2}$ & $\theta_{1}^{\rm ref}$ & $\theta_{2}^{\rm ref}$ & $\theta_{1}^{\infty}$ & $\theta_{2}^{\infty}$  \\ 
\hline \hline 
\multicolumn{13}{c}{$M = 50 M_{\odot}$} \\ 
\hline \hline 
1/8 & 0.5 & 0.95 & 1.40 & 1.67 \vline \vline & $49.9_{-1.2}^{+1.2}$ & $0.125_{-0.006}^{+0.006}$ & $0.50_{-0.01}^{+0.02}$ & $0.60_{-0.51}^{+0.34}$ & $1.41_{-0.05}^{+0.06}$ & $1.65_{-0.69}^{+0.59}$ & $1.42_{-0.05}^{+0.05}$ & $1.54_{-0.64}^{+0.65}$ \\ 
\hline 
1/8 & 0.95 & 0.5 & 1.28 & 1.72 \vline \vline & $50.1_{-1.2}^{+1.1}$ & $0.124_{-0.005}^{+0.006}$ & $0.94_{-0.02}^{+0.02}$ & $0.32_{-0.29}^{+0.45}$ & $1.27_{-0.03}^{+0.03}$ & $1.74_{-0.99}^{+0.83}$ & $1.28_{-0.03}^{+0.03}$ & $1.61_{-0.93}^{+0.91}$ \\ 
\hline 
1/3 & 0.5 & 0.95 & 1.32 & 1.76 \vline \vline & $50.13_{-0.82}^{+0.88}$ & $0.33_{-0.02}^{+0.02}$ & $0.48_{-0.03}^{+0.03}$ & $0.91_{-0.16}^{+0.07}$ & $1.28_{-0.11}^{+0.11}$ & $1.84_{-0.19}^{+0.20}$ & $1.37_{-0.11}^{+0.12}$ & $1.69_{-0.19}^{+0.20}$ \\ 
\hline 
1/3 & 0.95 & 0.5 & 1.18 & 1.80 \vline \vline & $50.11_{-0.47}^{+0.51}$ & $0.33_{-0.01}^{+0.01}$ & $0.93_{-0.02}^{+0.02}$ & $0.56_{-0.13}^{+0.13}$ & $1.16_{-0.03}^{+0.03}$ & $1.83_{-0.18}^{+0.17}$ & $1.22_{-0.04}^{+0.04}$ & $1.55_{-0.18}^{+0.17}$ \\ 
\hline 
$^{\dag}$1/1.1 & 0.5 & 0.95 & 2.19 & 1.65 \vline \vline & $49.99_{-0.18}^{+0.20}$ & $0.91_{-0.06}^{+0.08}$ & $0.55_{-0.17}^{+0.32}$ & $0.80_{-0.37}^{+0.16}$ & $1.91_{-0.27}^{+0.26}$ & $1.83_{-0.23}^{+0.29}$ & $1.05_{-0.61}^{+0.57}$ & $2.65_{-0.33}^{+0.31}$ \\ 
\hline 
1/1.1 & 0.5 & 0.95 & 0.39 & 2.74 \vline \vline & $49.96_{-0.13}^{+0.13}$ & $0.91_{-0.01}^{+0.01}$ & $0.54_{-0.07}^{+0.07}$ & $0.94_{-0.08}^{+0.04}$ & $0.47_{-0.26}^{+0.31}$ & $2.87_{-0.17}^{+0.15}$ & $1.89_{-0.06}^{+0.08}$ & $1.78_{-0.06}^{+0.04}$ \\ 
\hline 
$^{\dag}$1/1.1 & 0.95 & 0.5 & 1.40 & 0.73 \vline \vline & $49.92_{-0.11}^{+0.13}$ & $0.92_{-0.05}^{+0.06}$ & $0.73_{-0.25}^{+0.21}$ & $0.70_{-0.23}^{+0.24}$ & $1.34_{-0.19}^{+0.18}$ & $1.05_{-0.26}^{+0.23}$ & $0.53_{-0.27}^{+0.25}$ & $1.79_{-0.31}^{+0.62}$ \\ 
\hline 
1/1.1 & 0.95 & 0.5 & 0.37 & 2.81 \vline \vline & $49.93_{-0.09}^{+0.09}$ & $0.90_{-0.01}^{+0.01}$ & $0.94_{-0.05}^{+0.04}$ & $0.57_{-0.07}^{+0.09}$ & $0.23_{-0.14}^{+0.19}$ & $2.79_{-0.27}^{+0.19}$ & $1.30_{-0.03}^{+0.04}$ & $1.21_{-0.11}^{+0.09}$ \\ 
\hline 
\multicolumn{13}{c}{$M = 100 M_{\odot}$} \\ 
\hline \hline 
1/8 & 0.5 & 0.95 & 1.37 & 1.68 \vline \vline & $99.9_{-1.4}^{+1.4}$ & $0.125_{-0.003}^{+0.003}$ & $0.50_{-0.01}^{+0.01}$ & $0.83_{-0.21}^{+0.14}$ & $1.37_{-0.04}^{+0.04}$ & $1.77_{-0.27}^{+0.31}$ & $1.34_{-0.03}^{+0.04}$ & $1.89_{-0.27}^{+0.30}$ \\ 
\hline 
1/8 & 0.95 & 0.5 & 1.19 & 1.75 \vline \vline & $100.07_{-0.92}^{+0.91}$ & $0.125_{-0.002}^{+0.002}$ & $0.945_{-0.008}^{+0.008}$ & $0.53_{-0.28}^{+0.26}$ & $1.19_{-0.02}^{+0.02}$ & $1.78_{-0.34}^{+0.36}$ & $1.20_{-0.02}^{+0.02}$ & $1.55_{-0.32}^{+0.36}$ \\ 
\hline 
1/3 & 0.5 & 0.95 & 1.28 & 1.79 \vline \vline & $100.08_{-0.80}^{+0.81}$ & $0.330_{-0.009}^{+0.009}$ & $0.49_{-0.03}^{+0.03}$ & $0.91_{-0.12}^{+0.08}$ & $1.28_{-0.10}^{+0.08}$ & $1.83_{-0.18}^{+0.23}$ & $1.38_{-0.10}^{+0.08}$ & $1.66_{-0.17}^{+0.23}$ \\ 
\hline 
$^{\dag}$1/3 & 0.95 & 0.5 & 1.23 & 1.46 \vline \vline & $100.08_{-0.60}^{+0.60}$ & $0.330_{-0.007}^{+0.007}$ & $0.95_{-0.01}^{+0.01}$ & $0.50_{-0.11}^{+0.10}$ & $1.22_{-0.03}^{+0.03}$ & $1.60_{-0.23}^{+0.24}$ & $1.15_{-0.03}^{+0.03}$ & $1.98_{-0.23}^{+0.24}$ \\ 
\hline 
$^{\dag}$1/1.1 & 0.5 & 0.95 & 2.21 & 1.64 \vline \vline & $100.08_{-0.83}^{+0.85}$ & $0.90_{-0.03}^{+0.03}$ & $0.73_{-0.24}^{+0.24}$ & $0.66_{-0.34}^{+0.30}$ & $1.96_{-0.14}^{+0.24}$ & $1.70_{-0.12}^{+0.23}$ & $1.41_{-0.61}^{+0.28}$ & $2.54_{-0.26}^{+0.28}$ \\ 
\hline 
1/1.1 & 0.5 & 0.95 & 0.34 & 2.73 \vline \vline & $100.41_{-0.55}^{+0.46}$ & $0.905_{-0.010}^{+0.010}$ & $0.58_{-0.09}^{+0.06}$ & $0.92_{-0.05}^{+0.06}$ & $0.57_{-0.31}^{+0.30}$ & $2.74_{-0.08}^{+0.14}$ & $1.95_{-0.07}^{+0.07}$ & $1.68_{-0.05}^{+0.03}$ \\ 
\hline 
$^{\dag}$1/1.1 & 0.95 & 0.5 & 1.35 & 0.88 \vline \vline & $99.79_{-0.45}^{+0.44}$ & $0.91_{-0.03}^{+0.03}$ & $0.77_{-0.22}^{+0.19}$ & $0.68_{-0.21}^{+0.26}$ & $1.27_{-0.16}^{+0.10}$ & $1.14_{-0.21}^{+0.16}$ & $0.38_{-0.23}^{+0.22}$ & $1.94_{-0.35}^{+0.63}$ \\ 
\hline 
1/1.1 & 0.95 & 0.5 & 0.37 & 2.88 \vline \vline & $99.77_{-0.34}^{+0.34}$ & $0.915_{-0.008}^{+0.008}$ & $0.90_{-0.05}^{+0.04}$ & $0.51_{-0.05}^{+0.05}$ & $0.17_{-0.12}^{+0.22}$ & $2.87_{-0.30}^{+0.19}$ & $1.29_{-0.03}^{+0.04}$ & $1.16_{-0.10}^{+0.06}$ \\ 
\hline 
\multicolumn{13}{c}{$M = 200 M_{\odot}$} \\ 
\hline \hline 
1/8 & 0.5 & 0.95 & 1.37 & 1.68 \vline \vline & $200.2_{-3.3}^{+3.4}$ & $0.125_{-0.004}^{+0.004}$ & $0.50_{-0.02}^{+0.02}$ & $0.87_{-0.19}^{+0.11}$ & $1.37_{-0.05}^{+0.06}$ & $1.66_{-0.38}^{+0.34}$ & $1.40_{-0.05}^{+0.06}$ & $1.50_{-0.37}^{+0.34}$ \\ 
\hline 
1/8 & 0.95 & 0.5 & 1.37 & 1.68 \vline \vline & $199.5_{-1.2}^{+1.2}$ & $0.125_{-0.002}^{+0.002}$ & $0.947_{-0.008}^{+0.007}$ & $0.44_{-0.14}^{+0.14}$ & $1.37_{-0.02}^{+0.02}$ & $1.78_{-0.44}^{+0.41}$ & $1.38_{-0.02}^{+0.02}$ & $1.47_{-0.44}^{+0.41}$ \\ 
\hline 
$^{\dag}$1/3 & 0.5 & 0.95 & 1.41 & 1.58 \vline \vline & $199.9_{-1.3}^{+1.3}$ & $0.335_{-0.007}^{+0.008}$ & $0.51_{-0.02}^{+0.03}$ & $0.88_{-0.14}^{+0.10}$ & $1.45_{-0.07}^{+0.09}$ & $1.50_{-0.24}^{+0.16}$ & $1.33_{-0.07}^{+0.10}$ & $1.72_{-0.24}^{+0.17}$ \\ 
\hline 
$^{\dag}$1/3 & 0.95 & 0.5 & 1.15 & 1.43 \vline \vline & $200.07_{-0.99}^{+1.00}$ & $0.333_{-0.006}^{+0.006}$ & $0.95_{-0.01}^{+0.02}$ & $0.47_{-0.09}^{+0.11}$ & $1.15_{-0.03}^{+0.03}$ & $1.43_{-0.28}^{+0.20}$ & $1.08_{-0.04}^{+0.04}$ & $1.85_{-0.29}^{+0.20}$ \\ 
\hline 
$^{\dag}$1/1.1 & 0.5 & 0.95 & 2.25 & 1.63 \vline \vline & $200.1_{-1.3}^{+1.3}$ & $0.91_{-0.02}^{+0.02}$ & $0.75_{-0.26}^{+0.22}$ & $0.64_{-0.32}^{+0.31}$ & $1.99_{-0.11}^{+0.23}$ & $1.68_{-0.10}^{+0.17}$ & $1.43_{-0.64}^{+0.26}$ & $2.54_{-0.21}^{+0.24}$ \\ 
\hline 
1/1.1 & 0.5 & 0.95 & 0.29 & 2.73 \vline \vline & $200.3_{-1.3}^{+1.4}$ & $0.911_{-0.007}^{+0.007}$ & $0.52_{-0.03}^{+0.03}$ & $0.93_{-0.07}^{+0.05}$ & $0.25_{-0.13}^{+0.18}$ & $2.84_{-0.14}^{+0.20}$ & $1.93_{-0.05}^{+0.06}$ & $1.72_{-0.04}^{+0.04}$ \\ 
\hline 
$^{\dag}$1/1.1 & 0.95 & 0.5 & 1.38 & 0.80 \vline \vline & $199.8_{-1.1}^{+1.1}$ & $0.91_{-0.01}^{+0.01}$ & $0.84_{-0.28}^{+0.14}$ & $0.64_{-0.16}^{+0.29}$ & $1.34_{-0.13}^{+0.06}$ & $1.02_{-0.24}^{+0.20}$ & $0.49_{-0.23}^{+0.17}$ & $2.01_{-0.41}^{+0.44}$ \\ 
\hline 
1/1.1 & 0.95 & 0.5 & 0.37 & 2.88 \vline \vline & $200.2_{-1.1}^{+1.1}$ & $0.905_{-0.009}^{+0.008}$ & $0.93_{-0.05}^{+0.04}$ & $0.52_{-0.03}^{+0.04}$ & $0.28_{-0.20}^{+0.15}$ & $2.84_{-0.24}^{+0.20}$ & $1.29_{-0.05}^{+0.05}$ & $1.18_{-0.12}^{+0.11}$ \\ 
\hline 
\end{tabular} 

\caption{\label{table:params_uneqspins} Injected and recovered parameters for unequal-spin injections. Rows marked by a $(${\dag}$)$ have the injected value of $\phi_{12}$ set to $0$, while it is $\pi$ for the unmarked cases. 
}
\end{table*}

\bibliography{../tilts_pe_paper}

\end{document}